\documentclass[aps,superscriptaddress,twocolumn,floatfix,a4paper]{revtex4-2}

\usepackage[cp1250]{inputenc}
\usepackage[T1]{fontenc} 

\usepackage{times} 

\usepackage{epsfig}

\usepackage{bbm}

\usepackage{orcidlink} 

\usepackage{hyperref}
\hypersetup{colorlinks=true, citecolor=blue, urlcolor=blue, linkcolor=blue}

\usepackage{latexsym}
\usepackage{amsmath}
\usepackage{amssymb}
\usepackage{amsfonts}
\usepackage{amsthm}
\usepackage{mathrsfs}
\usepackage{color,verbatim,graphics}
\usepackage{psfrag}
\DeclareMathAlphabet{\mathrsfs}{U}{rsfs}{m}{n}
\DeclareMathAlphabet{\mathpzc}{OT1}{pzc}{m}{it}
\DeclareMathAlphabet{\matheus}{U}{eus}{m}{n}
\DeclareMathAlphabet{\mathbbold}{U}{bbold}{m}{n}

\newcommand{\bra}[1]{\langle #1|}
\newcommand{\ket}[1]{|#1\rangle}
\newcommand{\ketbra}[1]{| #1\rangle \langle #1|}
\newcommand{\braket}[2]{\langle #1 | #2 \rangle}
 \newcommand{\be}{\begin{equation}}
\newcommand{\ee}{\end{equation}} \newcommand{\eea}{\end{eqnarray}}
\newcommand{\bea}{\begin{eqnarray}} 
\newcommand{\va}[1]{\ensuremath{(\Delta#1)^2}}

\newcommand{\ex}[1]{\ensuremath{\langle{#1}\rangle}}
\newcommand{\exs}[1]{\ensuremath{\langle{#1}\rangle}}

\newcommand{\eins}{\mathbbm{1}} \renewcommand{\qed}{\ensuremath{\hfill
\blacksquare}}

\newcommand{\kommentar}[1]{}
\newcommand{\trace}{{\rm Tr}} 
 
\newcommand{\forget}[1]{}

\newcommand{\EQ}[1]{Eq.~\eqref{#1}} \newcommand{\EQS}[1]{Eqs.~\eqref{#1}}
\newcommand{\EQL}[1]{Equation~\eqref{#1}} \newcommand{\SEC}[1]{Sec.~\ref{#1}}
\newcommand{\FIG}[1]{Fig.~\ref{#1}} \newcommand{\REF}[1]{Ref.~\cite{#1}}
\newcommand{\REFS}[1]{Refs.~\cite{#1}}

\newcommand{\APP}[1]{Appendix~\ref{#1}}

\newcommand{\FQ}{\mathcal F_{Q}}

\begin{document}

\title{Iterative optimization in quantum metrology and entanglement theory \\ using semidefinite programming}

\author{\'Arp\'ad Luk\'acs\,\orcidlink{0000-0002-5737-1393}}
\email{lukacs.arpad@wigner.hu}
\affiliation{Department of Mathematical Sciences, Durham University,  Stockton Road, DH1 3LE Durham, United Kingdom}
\affiliation{Department of Theoretical Physics, University of the Basque Country UPV/EHU, P.~O. Box 644, E-48080 Bilbao, Spain}
\affiliation{HUN-REN Wigner Research Centre for Physics, P.~O. Box 49, H-1525 Budapest, Hungary}

\author{R\'obert Tr\'enyi\,\orcidlink{0000-0002-2839-0472}}
\email{robert.trenyi@ehu.eus}
\affiliation{Department of Theoretical Physics, University of the Basque Country UPV/EHU, P.~O. Box 644, E-48080 Bilbao, Spain}
\affiliation{EHU Quantum Center, University of the Basque Country UPV/EHU, 48080 Leioa, Spain}
\affiliation{Donostia International Physics Center DIPC, Paseo Manuel de Lardizabal 4, E-20018 San Sebasti\'an, Spain}
\affiliation{HUN-REN Wigner Research Centre for Physics, P.~O. Box 49, H-1525 Budapest, Hungary}
\affiliation{Department of Theoretical Physics, University of Szeged, Tisza L. krt. 84-86, H-6720 Szeged, Hungary}

\author{Tam\'as V\'ertesi\,\orcidlink{0000-0003-4437-9414}}
\email{tvertesi@atomki.hu}
\affiliation{HUN-REN Institute for Nuclear Research, P.~O. Box 51,
H-4001 Debrecen, Hungary}

\author{G\'eza T\'oth\,\orcidlink{0000-0002-9602-751X}}
\email[Corresponding author: ]{toth@alumni.nd.edu}
\homepage{http://www.gtoth.eu}
\affiliation{Department of Theoretical Physics, University of the Basque Country UPV/EHU, P.~O. Box 644, E-48080 Bilbao, Spain}
\affiliation{EHU Quantum Center, University of the Basque Country UPV/EHU, 48080 Leioa, Spain}
\affiliation{Donostia International Physics Center DIPC, Paseo Manuel de Lardizabal 4, E-20018 San Sebasti\'an, Spain}
\affiliation{IKERBASQUE, Basque Foundation for Science, E-48009 Bilbao, Spain}
\affiliation{HUN-REN Wigner Research Centre for Physics, P.~O. Box 49, H-1525 Budapest, Hungary}

\begin{abstract}
We discuss efficient methods to optimize the metrological performance over local Hamiltonians in a bipartite quantum system. For a given quantum state, our methods find the best local Hamiltonian for which  the state outperforms separable states the most from the point of view of quantum metrology. We show that this problem can be reduced to maximizing the quantum Fisher information over a certain set of Hamiltonians. We present the quantum Fisher information in a bilinear form and maximize it by an iterative see-saw (ISS) method, in which each step is based on semidefinite programming. We also solve the problem with the method of moments that works very well for smaller systems.  Our approach is one of the efficient methods that can be applied for an optimization of the unitary dynamics in quantum metrology, the other methods being, for example, machine learning, variational quantum circuits, or neural networks. The advantage of our method is the fast and robust convergence due to the simple mathematical structure of the approach. We also consider a number of other problems in quantum information theory that can be solved in a similar manner. For instance, we determine the bound entangled quantum states that maximally violate the Computable Cross Norm-Realignment  (CCNR) criterion.
\end{abstract}

\date{\today}

\maketitle

\section{Introduction}

Quantum entanglement is at the heart of quantum physics and several quantum information processing applications \cite{Horodecki2009Quantum,Guhne2009Entanglement,Friis2019,Horodecki2021Quantum}. It also plays a central role in quantum metrology. It is known that interparticle entanglement is needed to overcome the shot-noise limit of the precision in parameter estimation in a linear interferometer \cite{Pezze2009Entanglement}. It has also been shown that the larger the achieved precision in parameter estimation with a multiparticle system, the larger the depth of entanglement the state must possess \cite{Hyllus2012Fisher,Toth2012Multipartite}. However, not all entangled states are useful for metrology \cite{Hyllus2010Not}. Even bound entangled states, which are considered very weakly entangled, can be more useful for metrology than some distillable entangled states \cite{Czekaj2015Quantum,Toth2018Quantum,Pal2021Bound}.

At this point, several important questions arise. How can we find the best quantum state for metrology for a given setup? Perhaps, historically one of the first methods was the iterative see-saw (ISS) method applied in the quantum Fisher information picture \cite{Macieszczak2013Quantum_arxiv,Toth2018Quantum,Chabuda2020Tensor,Pal2021Bound}, as well as in the Bayesian one \cite{DemkowiczDobrzanksi2011Optimal,Macieszczak2014Bayesian,Jarzyna2015True}. There are also methods based on an optimization over the purifications \cite{Escher2011General,Demkowicz-Dobrzanski2012The,DemkowiczDobrzanski2014Using}. It is also  possible to use variational techniques for efficiently searching in the high-dimensional space of quantum states, which was also applied for a similar task \cite{Koczor2020Variational,Beckey2022Variational}. The variational search methods can even be tailored for particular physical systems, such as tweezer arrays as programmable quantum sensors \cite{Kaubruegger2019Variational}.
  
Another relevant question is how we could find the optimal Hamiltonian for a given quantum state. In the qubit case, it has been studied how to optimize the metrological performance for local Hamiltonians that are all Pauli spin matrices rotated by some unitary \cite{Hyllus2010Not}. In this case, the best metrological performance achievable by separable states is a constant independent of the Hamiltonian. Any state that has a better metrological performance than that must be entangled. 

The problem of the metrological performance of a bipartite state of qudits with a dimension larger than two is more complicated. In this case, the local Hamiltonians cannot be transformed into each other by local unitaries, and the best metrological performance achievable by separable states depends on the Hamiltonian. Thus, we introduce the metrological gain as a general way to characterize the quantum advantage of a system with arbitrary dimensions. It is defined as the ratio of the quantum Fisher information and the maximum of the quantum Fisher information over separable states maximized over local Hamiltonians \cite{Toth2020Activating}. The metrological gain is convex in the state, and it cannot decrease if additional copies or ancilla systems are added \cite{Toth2020Activating}. In the multiparticle case, the metrological gain is related to multiparticle entanglement, as it lower bounds the entanglement depth \cite{Trenyi2024Activation}. The optimization over the local Hamiltonians can also be carried out via an ISS method \cite{Toth2020Activating,Pal2021Bound,Trenyi2024Activation}. Thus, our method uses an ISS algorithm for optimizing the unitary dynamics rather than for optimizing the state. Note that instead of using the metrological gain, one can also consider the difference of the quantum Fisher information and its maximum for separable states \cite{Tan2021Fisher}, or even consider the difference with respect to a bound that depends on the state \cite{Yadin2018Operational,Morris2020Entanglement,Gessner2016Efficient,Gessner2017Resolution}.

In this paper, we examine the optimization problem above in a great depth. We elaborate the point that local Hamiltonians with meaningful constraints form a convex set. Thus, we have to maximize the metrological gain over a convex set of local Hamiltonians. We find several algorithms to maximize the metrological gain. The first method is based on maximizing a quadratic expression using a see-saw over a bilinear expression, and its computational costs are minimal. We present other algorithms that need quadratic programming with quadratic constraints and relaxations. Such methods work for small systems. We test the methods and compare them to each other. In the second half of the paper, we explore other important applications, where ISS methods can help with optimization. Namely, we list a number of similar problems that can be solved in an analogous way, such as maximization of norms over a convex set, maximizing the Wigner-Yanase skew information, looking for the extremal eigenvalue of Hermitian matrices, and also for looking for states that violate maximally the Computable Cross Norm-Realignment (CCNR) criterion \cite{Rudolph2005Further,Chen2003AMatrix}.

Our work is also related to recent efforts to solve various problems of numerical optimization in quantum metrology, including the combination of the tasks mentioned above, such as using an ISS for optimizing the quantum state and the unitary dynamics \cite{Toth2020Activating}. An ISS-based method has been used for an optimization over a general dynamics using its Choi-Jamio{\l}kowski representation \cite{Len2022Quantum}. An ISS-based method has also been used for optimizing over adaptive strategies, when  coherently probing several independent quantum channels \cite{Kurdzialek2025Quantum}. An optimization over the probe state and the dynamics has been realized experimentally in a programmable quantum sensor realized in trapped cold ions \cite{Marciniak2022Optimal}. For the identification of the optimal quantum metrological protocol in large systems, matrix product operators have been used \cite{Chabuda2020Tensor}. Optimization methods, including ISS, has been considered in estimating the spectral density of a stochastic signal field \cite{Gardner2025Stochastic}. Finally, numerical optimization in quantum metrology in various complex scenarios has been considered including multiparameter estimation \cite{Albarelli2019Evaluating,Gessner2020Multiparameter,DemkowiczDobrzanski2020Multi-parameter,Zhang2022QuanEstimation_PRR,Meyer2021AVariational,MacLellan2024End-to-end}, and there are even methods that are based on semidefinite programmingx~\cite{bavaresco2024designing}.

Variational methods used for the optimization of the probe state have already been mentioned. General optimization methods have also been applied for the optimization of the dynamics and other complex tasks. Optimization of the controls has been carried out using neural networks and reinforcement learning \cite{Xu2019Generalizable}. Machine learning approaches have been used in Bayesian parameter estimation \cite{Nolan2021Machine}. The advantageous aspects of ISS methods to general optimization methods are discussed in \REF{dulian2025qmetropythonoptimization}.

Our work has a relation to resource theories, in which the quantum Fisher
information can be used to identify quantum resources
\cite{Tan2021Fisher,Chitambar2019Quantum}. Indeed, quantum Fisher information
can be ideal to characterize quantum systems realized in Noisy
Intermediate-Scale Quantum applications \cite{Meyer2021Fisher}. For a resource
theory, we need to identify a convex set of free quantum states and a set of
free operations. In our case, the free states are the separable states and the
free operations include the local unitary operations. We can use the robustness
of metrological usefulness defined in \REF{Toth2018Quantum} to quantify our
resource. It asks the question, how much noise can be added to the state such
that it is not useful anymore metrologically. In this paper, we will examine
another measure, the metrological gain mentioned above.

Finally, our work is relevant for the field that attempts to find a quantum advantage in experiments \cite{Leibfried2004Toward,Napolitano2011Interaction-based,Riedel2010Atom-chip-based,Gross2010Nonlinear}. The bound for quantum Fisher information for separable states has already been used to detect metrologically useful multipartite entanglement in several experiments \cite{Lucke2011Twin,Krischek2011Useful,Strobel2014Fisher}.

Our paper is organized as follows.  In \SEC{sec:background}, we review the basic
notions of entanglement theory and quantum metrology.  In \SEC{sec:convsetH}, we
discuss, how our optimization problem can be viewed as a maximization over a
convex set of Hamiltonians. We review the notion of metrological gain. In
\SEC{sec:optgain}, we present various methods to optimize the quantum Fisher
information over a local Hamiltonian. In \SEC{sec:other}, we consider
optimization problems in quantum information that can be solved in a similar
way. For instance, we determine the bound entangled quantum states that
maximally violate the CCNR criterion for
entanglement detection \cite{Rudolph2005Further,Chen2003AMatrix}.

\section{Background}
\label{sec:background}

In this section, we review entanglement theory and quantum metrology.

\subsection{Entanglement theory}

In this section, we summarize basic notions of entanglement theory
\cite{Horodecki2009Quantum,Guhne2009Entanglement}. A bipartite quantum state is
called separable if it can be written as a mixture of product states as
\cite{Werner1989Quantum}
\be
\varrho_{\rm sep}=\sum_k p_k \varrho^{(k)}_1 \otimes \varrho^{(k)}_2,
\label{eq:sep}
\ee
where $p_k$ are probabilities. If a quantum state cannot be decomposed as in
\EQ{eq:sep}, then the state is entangled.

Deciding whether a quantum state is entangled is a hard task in general.
However, there are some necessary conditions for separability, that are easy to
test. If these conditions are violated then the state is entangled. 

One of the most important conditions of this type is the condition based on the
positivity of the partial transpose (PPT). For a bipartite density matrix given
as
\be
\varrho=\sum_{kl,mn}\varrho_{kl,mn} \ket{k}\bra{l}\otimes \ket{m}\bra{n}
\ee
the partial transpose according to first subsystem is defined by exchanging
subscripts $k$ and $l$ as 
\be
\varrho^{T1}=\sum_{kl,mn}\varrho_{lk,mn} \ket{k}\bra{l}\otimes \ket{m}\bra{n}.
\ee
It has been shown that for separable quantum states
\cite{Peres1996Separability,Horodecki1996Separability}
\be
\varrho^{T1}\ge 0
\ee
holds. Thus, if $\varrho^{T1}$ has a negative eigenvalue then the quantum state
is entangled. For $2\times2$ and $2\times3$ systems, the PPT condition detects
all entangled states \cite{Horodecki1996Separability}. For systems of size
$3\times3$ and larger, there are PPT entangled states
\cite{Horodecki1997Separability,Horodecki1998Mixed-State}. They are also quantum
states that are called bound entangled since their entanglement cannot be
distilled to singlet states with local operations and classical communication
(LOCC). Some of the bound entangled states are detected as entangled by the CCNR
criterion  \cite{Rudolph2005Further,Chen2003AMatrix}.

\subsection{Quantum metrology}
\label{sec:qmet}

In this section, we summarize basic notions of quantum metrology such as the quantum Fisher information and the Cram\'er-Rao bound. We will also discuss the multi-parameter case.

The fundamental task of quantum metrology is the following. We have a probe state or an initial state $\varrho,$ that evolves according to the dynamics
\be
\varrho_{\theta}=U_{\theta}\varrho U_{\theta}^\dagger,
\ee
where the unitary is given as
\be
U_{\theta}=\exp(-iH\theta).
\ee
Here $H$ is the Hamiltonian of the evolution, and $\theta$ is a parameter. Then, the metrological precision in estimating $\theta$ 
is bounded by the Cram\'er-Rao bound
\cite{Helstrom1976Quantum,Holevo1982Probabilistic,Braunstein1994Statistical,
Petz2008Quantum,Braunstein1996Generalized,Giovannetti2004Quantum-Enhanced,Demkowicz-Dobrzanski2014Quantum,Pezze2014Quantum,Toth2014Quantum,Pezze2018Quantum,Paris2009QUANTUM}
\be
\va \theta \ge \frac 1 {\nu \FQ[\varrho,H]},
\ee
where $\nu$ is the number of independent repetitions, and $\FQ[\varrho,H]$ is the quantum Fisher information,  defined as
\cite{Helstrom1976Quantum,Holevo1982Probabilistic,Braunstein1994Statistical,Petz2008Quantum,Braunstein1996Generalized}
\be
\FQ[\varrho,H]= \sum_{k,l}Q_{kl}^2 \vert H_{kl} \vert^2.
\label{eq:QFI}
\ee
Here, we consider a density matrix with the following eigendecomposition 
\be
\varrho=\sum_k \lambda_k |k\rangle\langle k|,\label{eigdecomp}
\ee
the constant coefficients $Q_{kl}$ depend only on the eigenvalues of the probe state, and are given as
\be
Q_{kl}:=\sqrt{2\frac{(\lambda_k-\lambda_l)^2}{\lambda_k+\lambda_l}},\label{eq:qqq}
\ee
whenever the denominator is not zero. The matrix elements of the Hamiltonian in the basis of the eigenvectors of
the density matrix are given as 
\be
H_{kl}=\langle k|H|l\rangle.
\ee
The quantum Fisher information in \EQ{eq:QFI} can be written in a different form that will be useful in our calculations as 
\bea
\FQ[\varrho,H]&=& \sum_{kl} Q_{kl}^2 \left[(H_{kl}^{\rm{r}})^2+
(H_{kl}^{\rm{i}})^2\right],
\label{eq:FQsq}
\eea
where  the superscripts r and i denote the real and imaginary parts,
respectively.

After defining the quantum Fisher information,
it is important to stress its basic properties,
namely, that it is convex in the state, and it is bounded from above by the variance as
\be
\FQ[\varrho,H]\le4\va{H}\label{eq:FQineq},
\ee
where for pure states there is an equality in \EQ{eq:FQineq}.

Moreover, there is an important inequality for the quantum Fisher information with the error propagation formula
\cite{Hotta2004Quantum,Escher2012Quantum_arxiv,Frowis2015Tighter}
\be
\va{\theta}_M=\frac{\va M}{\ex{i[M,H]}^2} \ge \frac 1 {\FQ[\varrho,H]},\label{eq:FQerrorprop}
\ee
which gives us, essentially, the precision of the $\theta$
parameter estimation, provided that we measure in the eigenbasis of the
observable $M.$ It can be shown that the uncertainty
$\va{\theta}_M$ is the smallest, and thus the bound in \EQ{eq:FQerrorprop} is
the best if we choose $M$ to be the so-called symmetric logarithmic derivative
\be
\label{eq:SLD}
M_{\rm opt}=2i\sum_{k,l}\frac{\lambda_{k}-\lambda_{l}}{\lambda_{k}+\lambda_{l}} \vert k
\rangle \langle l \vert \langle k \vert H \vert l \rangle,
\end{equation}
which fulfills
\be
i[\varrho,H]=\frac{1}{2}\{\varrho,M_{\rm opt}\},
\ee
where $\{X,Y\}=XY+YX$ is the anticommutator.
In this case, the inequality in \EQ{eq:FQerrorprop} is saturated. For pure states $\ket{\Psi},$ the symmetric logarithmic derivative is given as \cite{Toth2014Quantum}
\be
M_{{\rm opt},\Psi}=2i[\ketbra{\Psi},H].\label{eq:SLDpure}
\ee

So far we discussed the case of estimating a single parameter. Let us consider now the estimation of several parameters when the unitary dynamics is given by
\be
\exp\left(-i\sum_n H^{(n)} \theta^{(n)}\right),
\ee
where $H^{(n)}$ is the Hamiltonian corresponding to the parameter
$\theta^{(n)}.$ The Cram\'er-Rao bound can be generalized to this case as 
\begin{equation}\label{CR_mat}
C-\nu^{-1}F^{-1}\ge 0,
\end{equation}
where the inequality in \EQ{CR_mat} means that the left-hand side is a positive
semidefinite matrix, $C$ is now the covariance matrix with elements
\begin{equation}
C_{mn}=\exs{\theta^{(m)} \theta^{(n)}}-\exs{\theta^{(m)}}\exs{\theta^{(n)}},
\end{equation}
and $F$ is the Fisher matrix, and its elements are given as 
\bea
F_{mn}&\equiv& \FQ[\varrho,H^{(m)},H^{(n)}].
\eea
Here, the quantum Fisher information with two operators is given as 
\be
 \FQ[\varrho,A,B]=\sum_{k,l}Q_{kl}^2 A_{kl}B_{kl}^*\equiv \trace[(Q \circ Q
 \circ A) B], \label{eq:qFmat}
\ee
where $\circ$ denotes elementwise or Hadamard product. After straightforward algebra, the following equivalent formulation can be obtained 
\be
\FQ[\varrho,A,B]=\sum_{kl} Q_{kl}^2 \bigg[A_{kl}^{\rm{r}} B_{kl}^{\rm{r}} +
A_{kl}^{\rm{i}} B_{kl}^{\rm{i}}\bigg].
\label{eq:qFmat2}
\ee
If $\varrho$ has low rank, then, based on \EQ{eq:qqq}, the $Q$ matrix is sparse, which can greatly simplify the calculation of \EQ{eq:qFmat2}. In contrast to the single parameter case, the generalized Cram\'er-Rao bound in Eq.~\eqref{CR_mat} cannot be always saturated even in the limit of many repetitions.

\section{Basic notions for optimizing the metrological performance}
\label{sec:convsetH}

When maximizing the quantum metrological performance over a Hamiltonian, we have to consider two important issues.

\begin{figure}[h!]
  \includegraphics[scale=0.42]{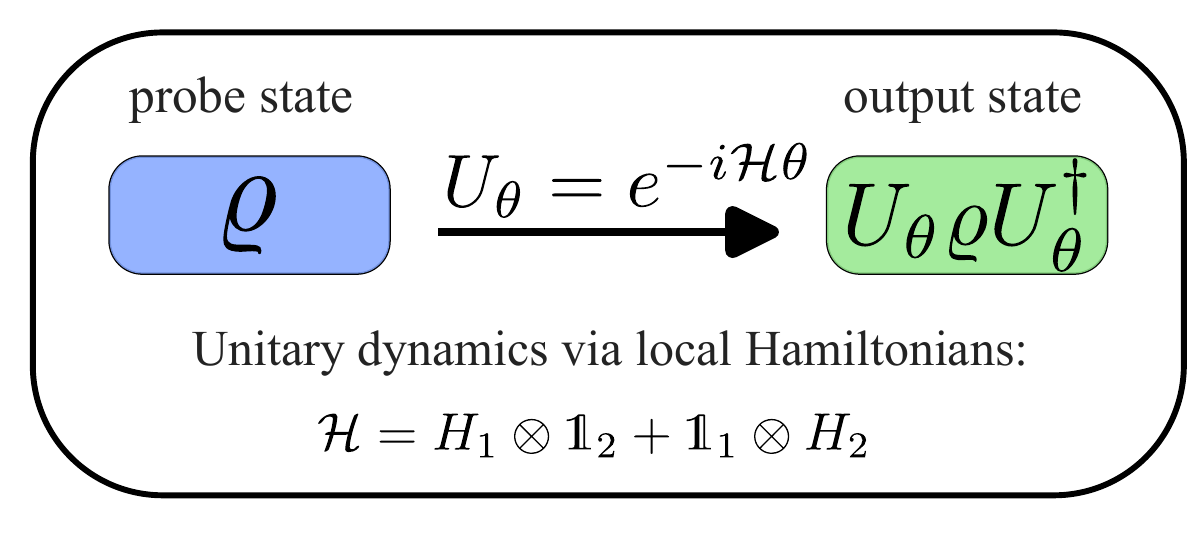}
  \caption{Scheme of the metrological process considered here. We start by
  preparing a given bipartite probe state $\varrho$ that undergoes the unitary
  dynamics $U_{\theta}=e^{-i\mathcal{H}\theta}$. The
 output state is $\varrho_{\theta}=U_{\theta}\varrho
  U_{\theta}^\dagger$, where $\theta$ is the parameter that we try to estimate.
  It is assumed that the Hamiltonian ${\mathcal H}$ that generates the unitary
  dynamics is local (i.~e., ${\mathcal H}\in{\mathcal L}$). A
  Hamiltonian ${\mathcal H}$ on a bipartite system is said to be local if it can
  be written as ${\mathcal H}=H_1\otimes \openone_2+\openone_1\otimes H_2,$
  where $H_1$ ($H_2$) is a Hamiltonian on the first (second) subsystem. Such Hamiltonians do not contain interaction terms between the two
  parties. In this setting, for a given probe state $\varrho$, we look for the
  optimal local Hamiltonian such that the output state 
  provides the best possible metrological precision
 in estimating the parameter $\theta.$
  We will discuss in the text, how to quantify the metrological 
  performance of a quantum state with a given local Hamiltonian.}
  \label{fig:proc_metrology}
\end{figure}

First, not all Hamiltonians are easy to realize. In particular, local Hamiltonians are much easier to realize with a high precision than Hamiltonians with an interaction term. Local Hamiltonians appear most typically in quantum metrology forming the basis of linear interferometers. Thus, we will restrict the optimization for such Hamiltonians given as
\begin{equation}
{\mathcal H}=H_1\otimes \openone_2+\openone_1\otimes H_2,\label{eq:bipartiteH}
\end{equation}
where $H_n$ are single-subsystem operators.  In the rest of the paper, ${\mathcal H}$ will denote such local Hamiltonians. Moreover, we will use $\mathcal L$ to denote the set of local Hamiltonians. With this, the considered metrological process is shown schematically in Fig.~\ref{fig:proc_metrology}.

Second, we have to note that for the quantum Fisher information 
\be
\FQ[\varrho,cH]=\vert c\vert ^2\FQ[\varrho,H]
\ee 
holds, where $c$ is a constant. Thus, it is easy to improve the metrological performance of any Hamiltonian by multiplying it by a constant factor. In order to obtain a meaningful optimization problem, we have to consider a formulation of such a problem where such a "trick" cannot be used to increase the quantum Fisher information to infinity. One option could be to divide the quantum Fisher information with the square of some norm or seminorm of the Hamiltonian, for
which
\be
\vert \vert c H \vert \vert= \vert c\vert \cdot \vert \vert H\vert \vert.
\ee
In principle, several norms for the normalization, for instance, the Hilbert-Schmidt norm 
\be
\vert\vert X\vert\vert_{\rm HS}=\sqrt{\trace(XX^\dagger)}\label{eq:HSnorm}
\ee
can be considered.  In this paper, we argue that it is advantageous to compare the metrological performance of a state to the best performance of separable states. Therefore, we will divide the quantum Fisher information of the state by the maximum quantum Fisher information achievable by separable states given as for some $\mathcal H\in\mathcal L$ as \cite{Ciampini2016Quantum,Toth2018Quantum}
\begin{equation} 
\FQ^{({\rm sep})}(\mathcal H)=\sum_{n=1,2} [ \sigma_{\max}(H_n)-\sigma_{\min}(H_n) ]^2,\label{eq:seplim}
\end{equation}
where $\sigma_{\max}(X)$ and  $\sigma_{\min}(X)$ denote the maximal and minimal eigenvalues, respectively. 

\subsection{Seminorm for local Hamiltonians}

Let us now present the details of our arguments. We will examine some properties of the maximum of the quantum Fisher information for separable states given in \EQ{eq:seplim}. Let us consider first the quantity 
\be
\delta(X)= \sigma_{\max}(X)-\sigma_{\min}(X),\label{eq:defdelta}
\ee
where $\sigma_{\min}(X)$ and $\sigma_{\max}(X)$ denote the smallest and largest
eigenvalues of $X,$ respectively. Simple algebra shows that
\cite{Boixo2007Generalized,*[{Inequalities similar to \EQ{eq:fqdelta}, also
involving the purity of the state and the variance, can be found at }] [{}]
{Toth2017Lower}}
\be
\FQ[\varrho,X]\le \delta^2(X). \label{eq:fqdelta}
\ee
With \EQ{eq:defdelta}, the maximum of the quantum Fisher information for
separable states, $\FQ^{({\rm sep})},$ can be expressed as \be\label{eq:FQsep2}
\FQ^{({\rm sep})}(\mathcal H)\equiv \delta^2(H_1)+\delta^2(H_2).
\ee

Next, let us prove some important properties of $\delta(X).$

{\bf Observation 1.} We know that $\delta(X)$ is a seminorm. While this is shown in \REF{Boixo2007Generalized}, for completeness and since the steps of the proof will be needed later, we present a brief argument to prove it.

{\it Proof.} We need the well-known fact that for Hermitian matrices $A$ and $B$  
\bea
\sigma_{\rm max}(A+B)&\le& \sigma_{\rm max}(A)+\sigma_{\rm max}(B),\nonumber\\
\sigma_{\rm min}(A+B)&\ge& \sigma_{\rm min}(A)+\sigma_{\rm
min}(B),\label{eq:sigmaminmax}
\eea
hold, following from Weyl's inequality \cite{Bhatia1997Matrix}. Based on these, we will now prove that $\delta(X)$ has all the properties of a seminorm \cite{ReedSimon1, RudinFA}, examining the three properties one by one. 

(i) Subadditivity or the triangle
inequality holds since 
\be
\delta(A+B)\le \delta(A)+\delta(B)
\ee 
for all $A,B$ based on  \EQ{eq:sigmaminmax}. 

(ii) Absolute homogeneity holds since \be \delta(cX)=|c|\delta(X)\label{eq:abshomdelta}\ee for all $c$ and $X.$ 

(iii) Non-negativity holds since $\delta(X)\ge 0$ holds for all $X.$ If $X$ is a zero matrix then $\delta(X)=0.$ However, $\delta(X)=0$ does not imply that $X$ is a zero matrix, thus $\delta(X)$  is not a norm, only a seminorm \cite{ReedSimon1, RudinFA}. $\qed$

Next, let us prove some important properties of the maximal quantum Fisher information for separable states.

{\bf Observation 2.} The expression with the square root of the maximum for separable states
\be
\sqrt{\FQ^{({\rm sep})}(\mathcal H)}
\ee
 is a seminorm for $\mathcal H\in\mathcal L.$ 

{\it Proof.} We will now follow steps similar to those in the proof of Observation 1, showing that the three properties of seminorms are satisfied.

(i) Let us consider a local Hamiltonian that is the sum of two local Hamiltonians
\be
\mathcal H=\mathcal H'+\mathcal H'',
\ee
where the two local Hamiltonians are given as 
\bea
{\mathcal H}'&=&H_1'\otimes \openone_2+\openone_1\otimes H_2',\nonumber\\
{\mathcal H}''&=&H_1''\otimes \openone_2+\openone_1\otimes H_2''.
\eea
Then, the following series of inequalities prove the subadditivity property

\bea
&&\sqrt{\delta^2(H_1'+H_1'')+\delta^2(H_2'+H_2'')}\nonumber\\
&&\quad\quad\le\sqrt{[\delta(H_1')+\delta(H_1'')]^2+[\delta(H_2')+\delta(H_2'')]^2}\nonumber\\
&&\quad\quad\le\sqrt{\delta^2(H_1')+\delta^2(H_2')}+\sqrt{\delta^2(H_1'')+\delta^2(H_2'')}.
\label{eq:ineqsubadd}
\eea
The first inequality in \EQ{eq:ineqsubadd} is due to the subadditivity of $\delta(X)$ and the monotonicity of $\sqrt{x^2+y^2}$ in $x$ and $y.$ The second inequality is based on the relation
\be
\sqrt{(x_1+x_2)^2+(y_1+y_2)^2}\le \sqrt{x_1^2+y_1^2}+\sqrt{x_2^2+y_2^2},
\ee
which can be proved by straightforward algebra.

(ii) Absolute homogeneity, 
\be
\sqrt{\FQ^{({\rm sep})}(c \mathcal H)} = |c|\sqrt{\FQ^{({\rm sep})}(\mathcal H)}
\ee 
for any real $c$ follows directly from the expression for $\FQ^{({\rm sep})}(\mathcal H)$  with $\delta(H_n)$ in Eq.~\eqref{eq:FQsep2}, using the absolute homogeneity property of $\delta$ given in
Eq~\eqref{eq:abshomdelta}

(iii) Non-negativity can be proven using  the expression for $\FQ^{({\rm sep})}(\mathcal H)$  with $\delta(H_n)$ in Eq.~\eqref{eq:FQsep2} and the non-negativity of $\delta$. $\qed$

Note that $\FQ^{({\rm sep})}(\mathcal H)=0$ holds if and only if $\mathcal
H=c\openone,$ where $c$ is a real constant. Thus, in a certain sense,
$\FQ^{({\rm sep})}(\mathcal H'-\mathcal H'')$ plays a role of distance between
local Hamiltonians $\mathcal H'$ and $\mathcal H''$ such that the distance is
zero if $\mathcal H'-\mathcal H''$ is the identity times some constant. In this
case, the two matrices are not identical, but from the point of view of quantum
metrology they act the same way.

Let us see some consequences of the observation. As any seminorm,
$\sqrt{\FQ^{({\rm sep})}(\mathcal H)}$ is convex in the local Hamiltonian. Due
to that, $\FQ^{({\rm sep})}(\mathcal H)$ is also convex.

\subsection{Metrological gain}

Based on these ideas, we define the metrological gain for a given local
Hamiltonian $\mathcal H\in\mathcal L$ and for a given quantum state $\varrho$ as
\cite{Toth2020Activating}
\be
g_{\mathcal H}(\varrho)=\frac{\FQ[\varrho,\mathcal H]}{\FQ^{({\rm
sep})}(\mathcal H)}.\label{eq:ggg}
\ee
In \EQ{eq:ggg}, we normalize the quantum Fisher information with a square of a
seminorm of local Hamiltonians. Then, we would like to obtain the gain based on an optimization over all local Hamiltonians as
\be
g(\varrho)=\max_{\mathcal H \in \mathcal L}g_{\mathcal H}(\varrho).
\ee
Note that $g(\varrho)$ is invariant under local unitaries $U_n,$ thus
\be
g(\varrho')=g(\varrho),
\ee
where the transformed state is defined as 
\be
\varrho'=(U_1 \otimes U_2) \varrho (U_1^\dagger \otimes U_2^\dagger).
\ee
Many entanglement measures, for instance, the Entanglement of Formation
has the above property \cite{Horodecki2009Quantum}.

We consider local Hamiltonians, as we have discussed at the beginning of \SEC{sec:convsetH}. While we do not study it in this work, it would also be interesting to consider non-local Hamiltonians for a similar definition. However, computing the analogous quantity for that case is difficult, since calculating the maximum of an operator expectation value for separable states is a hard task \cite{Horodecki2009Quantum}. Possibly, one can try to calculate the maximum for quantum states with a positive partial transpose (PPT) instead, which is an upper bound on the maximum for separable states, and it can be obtained via semidefinite programming (e.~g., \REFS{Toth2009Practical,Toth2018Quantum}).

Maximizing $g_{\mathcal H}$ over $\mathcal H$ is difficult since both the
numerator and denominator depend on $\mathcal H.$ Let us consider a subset of
Hamiltonians of the type \EQ{eq:bipartiteH} such that
\bea
\sigma_{\min}(H_n)=-c_n,\quad
\sigma_{\max}(H_n)=+c_n\label{eq:cc}
\eea
for $n=1,2.$ Then, the maximum of the quantum Fisher information for separable states is
\be
\FQ^{({\rm sep})}(\mathcal H)=4(c_1^2+c_2^2).
\ee
We can define the metrological gain for Hamiltonians that fulfill \EQ{eq:cc} as
\be
g_{c_1,c_2}(\varrho)=\max_{\mathcal H \in \mathcal L_{c_1,c_2}}
\frac{\FQ[\varrho,\mathcal H]}{4(c_1^2+c_2^2)},\label{eq:gc1c2B}
\ee
where we call $\mathcal L_{c_1,c_2}$ the set of local Hamiltonians satisfying
\EQ{eq:cc}. Finally, the maximal gain over all possible local Hamiltonians is given as 
\be
g(\varrho)=\max_{c_1,c_2}g_{c_1,c_2}.
\ee
Note that $g(\varrho)$ is convex in $\varrho$ \cite{Toth2020Activating}.

Based on \EQ{eq:gc1c2B}, we can see that computing $g_{c_1,c_2}(\varrho)$  is
the same as maximizing the quantum Fisher information for the given set of
Hamiltonians. We can do that by requiring that 
\begin{equation}
\label{eq:cnHconst}
c_n \eins \pm H_n \ge 0, 
\end{equation}
where $n=1,2$ and $c_n>0$ is some constant. This way we make sure that
\be
\sigma_{\min}(H_n)\ge -c_n, \quad \sigma_{\max}(H_n)\le+c_n,
\ee
for $n=1,2.$ Since we maximize a convex function over a convex set, the optimum
is taken on the boundary of the set where the eigenvalues of $H_n$ are $\pm
c_n.$ In this case, $H_n^2=c_n^2 \openone.$

\begin{figure}[t!]
\includegraphics[scale=1.9]{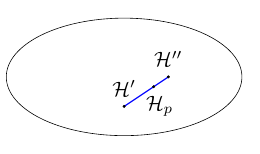}
\caption{The convex set of local Hamiltonians fulfilling  \EQ{eq:cnHconst}.
$\mathcal H_p$ is a "mixture" of $\mathcal H'$ and $\mathcal H''$ as given by \EQ{eq:Hp}. } \label{fig:convH}
\end{figure}

Note that the constraints in \EQ{eq:cnHconst} determine a convex set of
Hamiltonians. That is, if two Hamiltonians, $\mathcal H'$ and $\mathcal H''$,
are of the form \EQ{eq:bipartiteH} with the constraints given in
\EQ{eq:cnHconst}, then their convex combination, i.e., 
\be
\mathcal H_p=p\mathcal H'+ (1-p)\mathcal H''\label{eq:Hp}
\ee
with $0\le p\le1$ is also of that form, as shown in \FIG{fig:convH}. This can be
seen from \EQ{eq:sigmaminmax}.

Thus, we need to deal with convex sets, similarly, as convex sets appear in the
separability problem of quantum states
\cite{Werner1989Quantum,Horodecki2009Quantum}. In \APP{sec:specsep}, in order to
understand better the role of convexity in this case, we define a subset of
separable quantum states similar to the local Hamiltonians we consider in this
paper, and examine its properties.

In summary, we argue that the metrological gain, defined by \EQ{eq:ggg}, is
a natural way to characterize the metrological performance of quantum states as
it identifies states that perform better metrologically than all the separable
states. Moreover, it
considers local Hamiltonians, which are probably the least demanding to realize
experimentally. Finally, we also showed that the maximum of the quantum Fisher information for separable states is an important quantity.
Its square root is a seminorm over local Hamiltonians.  The maximum of the quantum Fisher information for separable states is used in the definition of the metrological gain to normalize the quantum Fisher information.

The metrological gain is also defined in the multiparticle case. It has been shown that, for the metrological gain the inequality \cite{Trenyi2024Activation}
\be
g(\varrho)\le k
\ee
holds, where $k$ is the entanglement depth. It is assumed that the local Hamiltonians have identical maximal and minimal eigenvalues, which is fulfilled in typical many-particle systems \cite{Trenyi2024Activation}. Thus, the advantage of using the metrological gain to characterize the performance of the setup is that it is directly related to multipartite entanglement. Moreover, maximal metrological gain is possible only with a maximal entanglement depth.

We remark that there are other quantities to characterize the
metrological performance.  One can consider the difference of the quantum Fisher
information and its maximum for separable states, and include the
possibility of non-unitary dynamics from the point of view of resource theories.
Considering the difference rather than the fraction makes the mathematical
treatment easier.
This way, all entangled states turn out to be a useful resource
\cite{Tan2021Fisher}. One can also consider the difference of the quantum Fisher
information and a bound that depends on the state, where the difference is
positive only for entangled states
\cite{Yadin2018Operational,Morris2020Entanglement,Gessner2016Efficient,Gessner2017Resolution}.
The advantage of this approach is that the quantity can be computed analytically.

\section{Method for maximizing the quantum Fisher information over local Hamiltonians}
\label{sec:optgain}

In the previous section, we have shown that maximizing the metrological gain is
essentially reduced to maximizing the quantum Fisher information with some
constraints on the local Hamiltonian given in \EQ{eq:cnHconst}, that is,
computing
\be
\max_{\mathcal H\in\mathcal L_{c_1,c_2}} \FQ[\varrho,\mathcal H],
\ee
where we remember that $\mathcal L_{c_1,c_2}$ is the set of local Hamiltonians
fulfilling \EQ{eq:cc}. Thus, in this section we will present methods that can
maximize the quantum Fisher information with such constraints.

\subsection{See-saw optimization based on a bilinear form}
\label{sec:Bilinear}

In \REF{Toth2020Activating}, a method was presented to maximize the quantum
Fisher information over local Hamiltonians for a given quantum state. It is an
iterative method, that needs to compute the symmetric logarithmic derivative
\EQ{eq:SLD} at each step and is based on quantum metrological considerations. In
this section, we will present a simpler formulation based on general ideas in
optimization theory. Later, this will also make it possible to find
provably the global maximum of the quantum Fisher information. 

The quantum Fisher information given in \EQ{eq:FQsq} is a weighted sum of
squares of the real and imaginary parts of the elements of the Hamiltonian. It can further be written as
\be
\vec v^T R \vec v,\label{eq:blinear}
\ee
where $R$ contains the $Q_{kl}$'s and $\vec v$ contains the $H_{kl}^{\rm{r}}$'s
and the $H_{kl}^{\rm{i}}$'s. The matrix $R$ is a positive semidefinite symmetric
matrix with real values. For the case of the quantum Fisher information, it is even a diagonal matrix. We will also consider other problems where it is a general  positive semidefinite symmetric matrix.
Minimizing such an expression over $\vec v$ is a problem that can efficiently be solved. In general, minimizing a convex function under constraints is a task appearing often in quantum metrology. For instance, there are efficient ways to obtain the minimum of the quantum Fisher information over a given set of states \cite{Apellaniz2017Optimal,Toth2015Evaluating,MullerRigat2023CertifyingQuantum}.
On the other hand, maximizing the expression given in \EQ{eq:blinear} over $\vec v$ is difficult. A usual way of maximizing such an expression is as follows
\cite{Konno1976Maximization,Konno1976Acutting,Floudas1995Quadratic}. We replace
the optimization over $\vec v$ by a simultaneous optimization of the vectors
$\vec v$ and $\vec w$ of a bilinear form as
\be
\max_{\vec v} \vec v^T R \vec v=\max_{\vec v,\vec w}  \vec v^T R \vec w. \label{eq:vw}
\ee
This can be a basis for a see-saw-type optimization. 

For completeness, let us prove \EQ{eq:vw} for the case of $R$ being a general  positive semidefinite  symmetric matrix. It is easy to see that 
the right-hand side is never smaller than the left-hand side. Now let us prove that the right-hand side is never larger than the left-hand side.
Let us start from
\be
(\vec v-\vec w)^T R(\vec v-\vec w)\ge 0. \label{eq:vw1}
\ee
Hence, follows that
\be
\vec v^T R \vec v+\vec w^T R \vec w\ge \vec v^T R \vec w+\vec w^T R \vec v=2\vec v^T R \vec w,\label{eq:vw2}
\ee
where the equality is due to the fact that $R$ is symmetric. Then, at least one of the following two inequalities hold
\bea
\vec v^T R \vec v &\ge& \vec v^T R \vec w,\nonumber\\
\vec w^T R \vec w &\ge& \vec v^T R \vec w.\label{eq:vw3}
\eea
After including an optimization over $\vec v$ and $\vec w$, the equality given in \EQ{eq:vw} follows. 

Based on the relation in \EQ{eq:vw}, we can write the following.

\begin{figure}[h!]
  \includegraphics[scale=0.5]{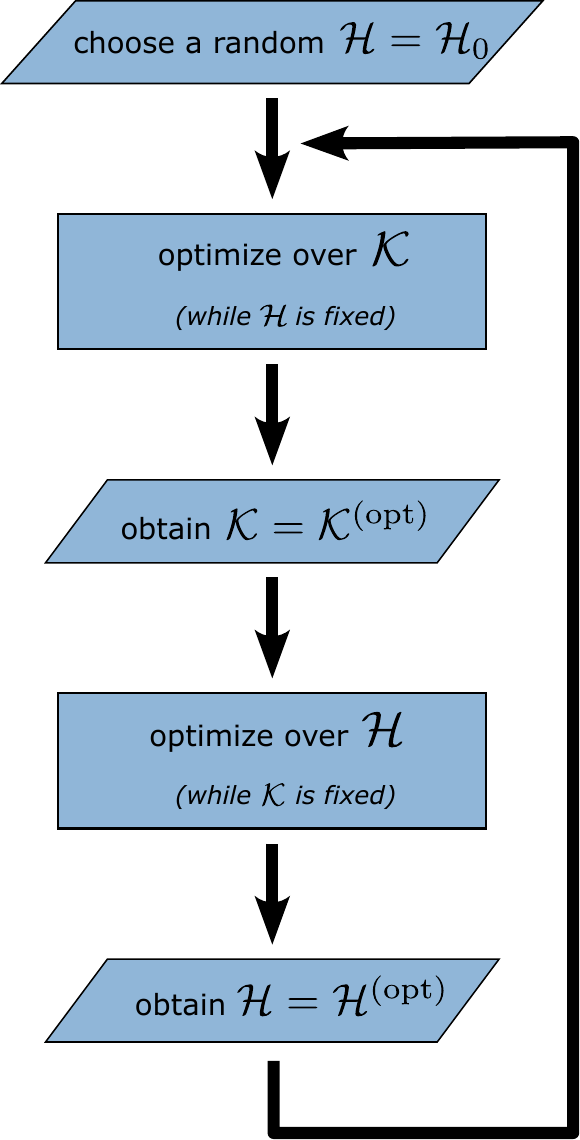}
  \caption{Based on the formula given in \EQ{eq:seesaw}, we can optimize the quantum
  Fisher information over local Hamiltonians for a fixed probe state with an
  iterative see-saw (ISS) method.} \label{fig:proc_ISS}
  \end{figure}

{\bf Observation 3.} The quantum Fisher information can be maximized with
an iterative see-saw (ISS) method. We
need to use the formula maximizing over local Hamiltonians
\bea
&&\max_{\mathcal H\in\mathcal L_{c_1,c_2}} \FQ[\varrho,\mathcal H]  \nonumber\\
&&\quad\quad=  \max_{ \mathcal  H,\mathcal K\in\mathcal L_{c_1,c_2}} \sum_{kl}
Q_{kl}^2 \bigg[\mathcal H_{kl}^{\rm{r}} \mathcal K_{kl}^{\rm{r}} + \mathcal
H_{kl}^{\rm{i}} \mathcal K_{kl}^{\rm{i}}\bigg]\nonumber\\
&&\quad\quad\equiv\max_{\mathcal H,\mathcal K\in\mathcal L_{c_1,c_2}}  \trace[(Q
\circ Q \circ \mathcal H) \mathcal K].
\label{eq:seesaw}
\eea
Based on \EQ{eq:seesaw}, we can carry out the following algorithm. Initially, we
choose $\mathcal H$ randomly. Then, we maximize over the
matrix $\mathcal K$ while keeping $\mathcal H$ fixed. Such an optimization can
be carried out, for instance, by semidefinite programming
\cite{Vandenberghe1996Semidefinite} or by simple matrix algebraic calculations,
see \APP{app:opt}. Then, we maximize over $\mathcal H$ while keeping $\mathcal
K$ fixed. Then, we optimize again over $ \mathcal K$, etc. 
This process is illustrated in Fig.~\ref{fig:proc_ISS}. Note that we do not
have to verify that the random initial Hamiltonian fulfills the constraints
given in \EQ{eq:cnHconst}, since a random initial Hamiltonian $\mathcal H$ and $c
\mathcal H$ for any $c>0$ as initial Hamiltonian will lead to the same optimal
$\mathcal K.$  (An alternative formulation of the optimization problem is in
\APP{app:alt}. With similar methods, we also analyze the optimization over a
quantum state rather than over the Hamiltonian \APP{app:state}.)

Based on \EQ{eq:qFmat2}, we can see that \EQ{eq:seesaw} can be expressed as
\be
\max_{\mathcal H\in\mathcal L_{c_1,c_2}} \FQ[\varrho,\mathcal H] =
\max_{\mathcal H,\mathcal K\in\mathcal L_{c_1,c_2}} \FQ[\varrho,\mathcal
H,\mathcal K].
\ee
This gives a meaningful physical interpretation of the see-saw method. 

Since $\FQ[\varrho,H]$ is convex in $H_{kl}^{\rm{r}}$ and $H_{kl}^{\rm{i}},$ if
we maximize it over a convex set of Hamiltonians, then it will take the maximum
at the boundary of the set. This optimization is a hard task, as we have discussed. We have to start
from several random initial conditions, and only in some of the cases we will find a
solution. 

\newlength{\mywidtha}
\settowidth{\mywidtha}{13}
\newlength{\mywidthb}
\settowidth{\mywidthb}{13.6421}
\newcommand{\minuswithspace}{\makebox[\mywidthb][c]{-}}

\begin{table*}[t!]
\begin{tabular}{l|llll|lll}
\hline
\hline
                                                                  &
                                                                  \multicolumn{4}{c|}{$\mathop{\rm
                                                                  max}F_Q[\varrho,
                                                                  H]$}  &
                                                                  \multicolumn{3}{c}{$\mathop{\rm
                                                                  max}4
                                                                  I_\varrho$}\\
\hline
Quantum state                                                     & Dim. &
See-saw & Level $1$ & Level $2$    & See-saw & Level $1$ & Level $2$\\
\hline
$\varrho_{AB}^{(0.1,2)}$  from \EQ{eq:iso}                  & 2    &
\makebox[\mywidtha][r]{13}.6421 &  \makebox[\mywidtha][r]{13}.6421   &
\makebox[\mywidtha][r]{13}.6421      & \makebox[\mywidtha][r]{10}.3338 &
\makebox[\mywidtha][r]{10}.3338   & \makebox[\mywidtha][r]{10}.3338  \\ 
$\varrho_{AB}^{(0.1,3)} \otimes \varrho_{A'B'}^{(0.1,3)}$      & 9    &
\makebox[\mywidtha][r]{13}.8828 &  \makebox[\mywidtha][r]{14}.0568   &
\minuswithspace     &  \makebox[\mywidtha][r]{13}.4330 &
\makebox[\mywidtha][r]{13}.6009  & \minuswithspace \\
Horodecki, $a=0.3$  \cite{Horodecki1998Mixed-State}               & 3    &
\makebox[\mywidtha][r]{4}.8859 &   \makebox[\mywidtha][r]{5}.1280   &
\makebox[\mywidtha][r]{4}.8859      &  \makebox[\mywidtha][r]{4}.8020 &
\makebox[\mywidtha][r]{4}.9910   &  \makebox[\mywidtha][r]{4}.8020  \\ 
UPB \cite{Bennett1999Unextendible,DiVincenzo2003Unextendible}     & 3    &
\makebox[\mywidtha][r]{5}.6913 &  \makebox[\mywidtha][r]{6}.1667   &
\makebox[\mywidtha][r]{5}.6913      &  \makebox[\mywidtha][r]{5}.6913 &
\makebox[\mywidtha][r]{6}.1667   &  \makebox[\mywidtha][r]{5}.6913  \\ 
Met. useful \ PPT \cite{Toth2018Quantum}                                  & 4
&  \makebox[\mywidtha][r]{9}.3726 &  \makebox[\mywidtha][r]{9}.3726   &
\minuswithspace           &  \makebox[\mywidtha][r]{9}.3726 &
\makebox[\mywidtha][r]{9}.3726   &  \minuswithspace       \\ 
Private PPT \cite{Badziag2014Bound_with_proper_ogonek,Pal2021Bound} & 6    &
\makebox[\mywidtha][r]{10}.1436 & \makebox[\mywidtha][r]{10}.1436   &
\minuswithspace           & \makebox[\mywidtha][r]{10}.1436 &
\makebox[\mywidtha][r]{10}.1436   &  \minuswithspace       \\ 
\hline
\hline
\end{tabular}
\caption{Maximum of the quantum Fisher information and the Wigner-Yanase skew
information with the see-saw method for $c_1=c_2=1$ for various quantum states
described in the text. The results for quadratic programming for various levels
of approximation are also shown. "UPB" refers to an Unextendible
Product Basis state discussed in the main text \cite{Bennett1999Unextendible,DiVincenzo2003Unextendible}. "Met. useful PPT" denotes the metrologically
useful PPT state considered in Ref.\ \cite{Toth2018Quantum}, whereas "Private
PPT" denotes the PPT state based on private states given in Refs.\
\cite{Badziag2014Bound_with_proper_ogonek,Pal2021Bound}.  }
\label{tab:QuadraticAndWY}
\end{table*}

We tested our method, together with the method of \REF{Toth2020Activating}. We
used MATLAB \cite{MATLAB2020} and semidefinite programming using MOSEK
\cite{MOSEK} and the YALMIP package \cite{Lofberg2004Yalmip}. We also used the
QUBIT4MATLAB package \cite{Toth2008QUBIT4MATLAB,QUBIT4MATLAB_actual_note_href}.

We tested the method for a number of quantum states. We considered the isotropic
state of two $d$-dimensional qudits defined as
\be
\varrho_{AB}^{(p,d)}=(1-p)\ketbra{\Psi_{\rm
me}}+p\frac{\openone}{d^2},\label{eq:iso}
\ee
where $\ket{\Psi_{\rm me}}$ is the maximally entangled state and $p$ is the
fraction of the white noise. In particular, we considered $p=0.1$ and $d=2$. For
such states, the optimal Hamiltonian is known analytically
\cite{Toth2020Activating}. We also considered two copies of the state with $p=0.1$
and $d=3$ given as $\varrho_{AB}^{(0.1,3)} \otimes \varrho_{A'B'}^{(0.1,3)},$
where the bipartite system consists of the parties $AA'$ and $BB'.$ We also
considered some bound entangled states that have been examined in the
literature, and are often used to test various algorithms. Such states have a
nontrivial structure, are highly mixed, but can be metrologically useful. We
looked at the $3\times 3$ state based on an Unextendible
Product Basis (UPB) \cite{Bennett1999Unextendible,DiVincenzo2003Unextendible},
the $3\times 3$ Horodecki state for $a=0.3$ \cite{Horodecki1998Mixed-State}, the
$4\times 4$ metrologically useful bound entangled state in
\REF{Toth2018Quantum}, and the $6\times 6$ private state given in
\REF{Badziag2014Bound_with_proper_ogonek,Pal2021Bound}.  The results are shown
in Table~\ref{tab:QuadraticAndWY}.  The success probability of the method of
\REF{Toth2020Activating} is similar, however, the formulation used in this paper
makes it possible to design methods that provably find the optimum \footnote{We
started from a random initial Hermitian Hamiltonian, where the real and
imaginary parts of the elements had a normal distribution with a zero mean and a
unit variance, and they were independent from each other, apart from the
requirement of the Hermiticity.}.

\subsection{Quadratic programming with quadratic constraints}
\label{sec:Quadprog}

In the previous section, we considered methods that aim to maximize the quantum
Fisher information over the Hamiltonian numerically. In practice, they are very
efficient. Still, it is not certain that they find the global optimum and they
might find a smaller value. In this section, we present methods that always find
the global optimum. However, they work only for small systems. We also present
simple approximations of such methods called relaxation. They find either the
maximum or a value larger than that. The method given in \SEC{sec:Bilinear}
looking for the maximum with the see-saw and the methods discussed in this
section can be used together. The gap between the two results gives information
about the location of the maximum. If the two methods find the same value, then
it is the global maximum.

The basic idea of the method of moments can be understood as follows. According
to the Shor relaxation  \cite{Burer2020Exact,*[{}] [{. Originally published in
Sov. J. Comput. Syst. Sci. {\bf 25}, 1-11 (1987).}] Shor1987Quadratic}, the
maximization of \EQ{eq:blinear} is replaced by the maximization of
\be
{\rm Tr}(R X),\label{eq:trRX}
\ee
where the Hermitian $X$ is constrained as 
\be
X \ge \vec v \vec v^T.\label{eq:const}
\ee
Note that \EQ{eq:const} can easily be coded in semidefinite programming. We can
also add further and further conditions on $X$ that lead to lower and lower
values for the maximum. Similar relaxation techniques have been used in
entanglement theory to detect entanglement,
non-locality~\cite{Navascues_2008}, to obtain the maximum of
an operator for separable states  \cite{Eisert2004Complete}, or
even to determine the ground state energy in spin one-half Heisenberg
models~\cite{Wang2024Certifying}.

For our problem, we used the
method of moments \cite{Lasserre2001Global}. 

{\bf Observation 4.} The maximization of quantum Fisher information can be
carried out with the method of moments. In this case, the conditions of
\EQ{eq:cnHconst} are replaced by
\begin{equation}
\label{eq:cnHconst2}
H_n^2  = c_n^2 \openone,
\end{equation}
for $n=1,2$, which makes sure that all eigenvalues of $H_n$ are $\pm c_n.$ The
results of the first level relation can be improved if we add the constraints
\begin{equation}
{\rm Tr}(H_n)=m_nc_n
\end{equation}
for $n=1,2,$ where $m_n=-d,-d+2,...,d-2,d,$ where this way we set the number of $+c_n$'s  to a given value in the diagonal of $H_n$. In quadratic programming, we can
also use the constraint 
\be
H_1^2+H_2^2\le \openone. \label{eq:const2}
\ee
After simple considerations, we can find that \EQ{eq:const2}  is equivalent to  \EQ{eq:cnHconst2} for all $c_n$ values. Thus, we do not
need to carry out calculations for a range of $c_n$ values. A somewhat weaker
constraint, which is computationally simpler is
\be
{\rm Tr}(H_1^2+H_2^2)\le 1.
\ee

We tested our methods for the same states we considered in \SEC{sec:Bilinear}.
For the $2\times2$ example, on a laptop computer, we could carry out
calculations at level $1, 2$ and $3.$ Even level $1$ gave the exact result. For
the $3\times3$ examples, we could carry out  calculations at level $1$ and $2,$
and level $2$ gave the correct result. For the $4\times4$ and $6\times6$
examples, the level $1$ calculation gave already the correct result. The results
are shown in  Table~\ref{tab:QuadraticAndWY}.
A detailed explanation of the method via a concrete example is in \APP{app:FQ_MOMENT_EXAMPLE}.

\section{Other problems in which sums of quadratic expressions must be maximized over a convex set}
\label{sec:other}

In this section, we consider optimization problems in quantum information
science that can be formulated similarly to the maximization of the quantum
Fisher information.

\subsection{Optimization over various subsystems alternatingly}

In quantum information, see-saw methods have already been used
\cite{Werner2001Bell,Werner2001All,Liang2007Bounds,Pal2010Maximal,Navascues2020Connector}.
Problems related to entanglement in multipartite systems seem to be ideal for
such methods. For instance, a typical problem is maximizing an operator
expectation value for a product state
\be
\varrho=\varrho_1 \otimes \varrho_2.
\ee
Then, we can maximize alternatingly over $\varrho_1$ and over $\varrho_2$.

\subsection{Maximization of the Wigner-Yanase skew information}

The Wigner-Yanase skew information is defined as \cite{Wigner1963INFORMATION}
\be
I_\varrho(\mathcal H)={\rm Tr}(\mathcal H^2\varrho)-{\rm Tr}(\mathcal
H\sqrt{\varrho} \mathcal H\sqrt{\varrho}).\label{eq:IWY}
\ee
Here, we optimize over $\mathcal H.$  It is clear how to maximize \EQ{eq:IWY}
with a see-saw method.  
\be
\max_{\mathcal H,\mathcal K\in\mathcal L_{c_1,c_2}} I_\varrho(\mathcal H)={\rm Tr}\left[\frac 1 2 (\mathcal H \mathcal K + \mathcal K \mathcal H)\varrho\right]-
{\rm Tr}(\mathcal
H\sqrt{\varrho} \mathcal K\sqrt{\varrho}).
\ee
Here, instead of $\mathcal H \mathcal K$ we use $\frac 1 2 (\mathcal H \mathcal K + \mathcal K \mathcal H),$ since the latter is Hermitian. The expectation value of a non-Hermitian matrix could be complex.
Then, the method finds the maximum based on the derivation presented in \EQS{eq:vw1}, \eqref{eq:vw2}, and \eqref{eq:vw3}.
Maximizing the Wigner-Yanase skew information can be used
to lower bound the quantum Fisher information based on  \cite{Wigner1963INFORMATION}
\be
\FQ[\varrho,\mathcal H]\ge 4 I_\varrho(\mathcal H).\label{eq:boundWY}
\ee
The results are shown in Table~\ref{tab:QuadraticAndWY}. From a comparison with
the results for the quantum Fisher information, we can see that for the last
three of the five examples, there is an equality in \EQ{eq:boundWY}. This has
been known for the last two examples \cite{Toth2018Quantum,Pal2021Bound}. For
these states, we also have an equality in \EQ{eq:FQineq} with the optimal
Hamiltonian, that is, the quantum Fisher information equals four times the variance. Surprisingly, for the fourth example, the UPB state, we do not have
an equality in \EQ{eq:FQineq}, which we can check with direct calculation. 

\subsection{Searching for the eigenstate with maximal/minimal
eigenvalue}
\label{eq:poweriteration}
  
In this section, we discuss how to look for the largest eigenvalue of a positive semidefinite Hermitian matrix
with a see-saw method similar to the one discussed in \SEC{sec:Bilinear}.
Interestingly, the see-saw is guaranteed to converge in this case.

We need the relation that converts the maximization over a quadratic form into a
maximization over a bilinear expression as 
\be
\max_{\ket{\Psi}} \ex{\Psi\vert A\vert \Psi}=\max_{\ket{\Psi},\ket{\Phi}}
\ex{\Psi\vert A\vert \Phi},\label{eq:maxphi}
\ee
where $A$ is a Hermitian matrix and the right-hand side of \EQ{eq:maxphi} can
immediately be used to define a see-saw algorithm.

Let us now consider the Cauchy-Schwarz inequality
\be\label{eq:inequa} \ex{\Psi \vert A \Phi} \le \sqrt{\ex{A \Phi \vert A \Phi}
\ex{ \Psi \vert \Psi}}=\sqrt{\ex{A \Phi \vert A \Phi}},
\ee
where we took into account that $\ex{ \Psi \vert \Psi}=1$. The inequality is
saturated for 
\be
\ket\Psi= \ket{A \Phi} / \sqrt{\ex{A \Phi \vert A \Phi}}.\label{eq:psi1step}
\ee 

Based on these considerations, the see-saw procedure is given as follows.

(i) Choose some random $\ket\Phi$ unit
vector.

(ii) Maximize over $\ket\Psi$  and $\ket\Phi$ alternatingly. First, $\ket\Psi$ is chosen based on \EQ{eq:psi1step}.
Then, in the alternate step, we use 
\be
\ket\Phi = \ket{A\Psi}/ \sqrt{\ex{A \Psi \vert A \Psi}}.
\ee

With these steps, we realized the power method for finding
the largest eigenvalue  of a positive self-adjoint operator, which is a classic iterative algorithm in numerical linear algebra \cite{Saad2011Numerical,Golub2013Matrix}. 
The iteration is known to
converge as long as the starting vector ($\ket\Phi$ in our case) overlaps with
the eigenvector corresponding to the maximal eigenvalue of $A$.
If $A$ is not positive semidefinite, then the method converges to the eigenvalue with the largest absolute value. Note that the problem could have also been solved with a one shot semidefinite program without iteration.

 \subsection{Maximization of the Hilbert-Schmidt norm}
 
In this section, we look at the problem of maximizing a matrix norm with a method similar to the
one discussed in \SEC{sec:Bilinear}. Let us consider the Hilbert-Schmidt norm
given in \EQ{eq:HSnorm}, for which the maximization can be rewritten
\be
\max_{X\in S} \vert\vert X \vert\vert_{\rm HS}^2 =\max_{X,Y\in S}{\rm
Tr}\left(XY^\dagger\right),\label{eq:hs}
\ee
where $S$ is some set of matrices. A straightforward see-saw optimization can be
formulated based on \EQ{eq:hs}. If $S$ is the set of density matrices then the
maximization of the Hilbert-Schmidt norm is just the maximization of the purity
${\rm Tr}(\varrho^2)$ or the minimization of the linear entropy $S_{\rm
lin}(\varrho)=1-{\rm Tr}(\varrho^2).$

For instance, we can consider obtaining
\bea
&&\text{$\max_{
{\scriptsize\begin{array}{l}\varrho:\;\trace(O^{(k)}\varrho)=e^{(k)}, \\ k=1,2,...,M\end{array}}
}$}\; \trace(\varrho^2), 
\eea
where $\varrho$ is a density matrix, $O^{(k)}$ are operators, and $e^{(k)}$ are
constants. The constraints require that the expectation values of the  $O^{(k)}$ operators equal the constants $e^{(k)}.$
The optimization problem will lead to pure states if there are pure
states satisfying the condition.
The optimization can be solved with the see-saw method as follows
\bea
&&\text{$\max_{
{\scriptsize\begin{array}{l}\varrho_1,\varrho_2:\;\trace(O^{(k)}\varrho_n)=e^{(k)}, \\ n=1,2, \;k=1,2,...,M\end{array}}
}$}\; \trace(\varrho_1 \varrho_2), 
\eea
where $\varrho_1$ and $\varrho_2$ are density matrices. 
The see-saw method shows a rapid convergence.
In fact, it converges typically in a single step.  
The problem can also be solved with the method of moments.  
After examining this simple
case, in addition to the linear constraints, we can have more complicated constraints on the quantum state. For
instance, we can consider PPT states or states with a given negativity
\cite{Toth2018Quantum}.

We add that in entanglement theory, the optimization over pure product states of bipartite and
multipartite systems is an important task. This can be carried out if we require
that the reduced density matrices are pure \cite{Eisert2004Complete}. In this
case,  the purity of the reduced states $\varrho_n$ appears among the
constraints as
\be
 \trace(\varrho_n^2) =1.
\ee
We add that optimization over pure states and rank constrained optimization is
possible with other methods using semidefinite programming
\cite{Gross2010Quantum,Yu2022Quantum-Inspired}.

 \subsection{Maximization of the trace norm}

Another important norm in quantum physics, the trace norm is defined as
\be
\vert\vert X \vert\vert_{\rm tr} ={\rm
Tr}\left(\sqrt{XX^\dagger}\right).\label{eq:trnorm}
\ee

{\bf Observation 5.} The trace norm of a Hermitian matrix given in
\EQ{eq:trnorm} can be maximized with a see-saw method.

{\it Proof.} In order to obtain an optimization over a bilinear form, we use the
dual relation between the trace norm and the spectral (operator) norm
\cite{HiaiPetz, Bhatia1997Matrix}, based on which
\be
\vert\vert X \vert\vert_{\rm tr}=\max_{Y\in {\mathbbm C}^{m\times m}:YY^{\dagger}\le
\openone} {\rm Tr}(X^{\dagger}Y).\label{eq:trdual}
\ee
The maximization of $\vert\vert X \vert\vert_{\rm tr}$ over a set $S$ can then
be written as 
\be
\max_{X\in S}\vert\vert X \vert\vert_{\rm tr}=\max_{X\in S}\max_{Y\in 
{\mathbbm C}^{m\times m}:YY^{\dagger}\le \openone} {\rm Tr}(X^{\dagger}Y),\label{eq:trdual2}
\ee
which can be calculated by a see-saw similar to the ones discussed before. We
have to maximize ${\rm Tr}(X^{\dagger}Y)$ alternatingly by $X$ and $Y.$ $\qed$

In entanglement theory, the trace norm is used to define the CCNR criterion  for
entanglement detection \cite{Rudolph2005Further,Chen2003AMatrix}. The criterion says that
for every bipartite separable state $\varrho,$ i.e., states of the form given in
\EQ{eq:sep},  we have
\be
\vert\vert R(\varrho) \vert\vert_{\rm tr}\le 1,\label{eq:ccnr}
\ee
where $R(\varrho)$ is the realigned matrix obtained by a certain permutation of
the elements of $\varrho.$ If \EQ{eq:ccnr} is violated then the state is
entangled. Clearly, the larger the left-hand side of \EQ{eq:ccnr}, the larger the
violation of \EQ{eq:ccnr}.

The CCNR criterion can detect weakly entangled quantum states, called bound
entangled states, that have a positive partial transpose. From this point of
view, it is a very relevant question to search for PPT states which violate
\EQ{eq:ccnr} the most. This can be done by maximizing $\vert\vert R(\varrho)
\vert\vert_{\rm tr}$ over PPT states. We carried out a numerical optimization
using semidefinite programming (e.~g., optimization over PPT states has been considered in \REFS{Toth2009Practical,Toth2018Quantum}). We present the maximum of $\vert\vert R(\varrho)
\vert\vert_{\rm tr}$ for PPT states for various dimensions in
Table~\ref{tab:CCNR}. The state for the $4\times 4$ case could be determined
analytically, see \APP{app:BES}. 

\begin{table}[t!]
\begin{tabular}{l|llll}
\begin{tabular}{c} Dimension\\ $d_1\times d_2$ \end{tabular}&\begin{tabular}{c} Maximum of
\\ $\vert\vert R(\varrho) \vert\vert_{\rm tr}$\end{tabular} \\
\hline
$2\times2$&              1\\
$2\times4$&              1\\
$3\times3$&              1.1891 \\
$3\times4$&              1.2239 \\
$4\times4$&              1.5 \\
$5\times5$&              1.5 \\
$6\times6$&              1.5881\\
\end{tabular}
\caption{The largest values for $\vert\vert R(\varrho) \vert\vert_{\rm tr}$ for
$d_1\times d_2$ PPT states for various $d_1$ and $d_2.$ The results were obtained by numerical
maximization. } \label{tab:CCNR}
\end{table}

\section{Conclusions}\label{sec:conc}

We discussed how to maximize the metrological gain over local Hamiltonians for a bipartite quantum system. We presented methods that exploit the fact that the quantum Fisher information can be written as a bilinear expression of the elements of the Hamiltonian, if the Hamiltonian is given in the basis of the density matrix eigenstates. We showed an iterative see-saw (ISS) method that can maximize the quantum Fisher information for large systems. We showed another method based on quadratic programming. With this so-called relaxation method we can confirm whether the maximum found by the ISS method is indeed a global one for small systems.

We tested the two approaches on concrete examples, which include one and two copies of the isotropic state of two $d$-dimensional qudits and some important bound entangled states, like, for instance, the $3\times3$ Horodecki state. These examples also show that we have provided a simple, systematic and efficient method for maximizing the metrological performance (i.e., the metrological gain) of bipartite quantum states with local Hamiltonians. Our work can straightforwardly be generalized to multiparticle systems. 

We also considered several similar problems that can be solved based on such ideas, for instance, obtaining the maximum violation of the Computable Cross Norm-Realignment  (CCNR)  criterion for states with a positive partial transpose, maximizing various norms, searching for the extremal eigenvalues of Hermitian matrices, and also maximizing the Wigner-Yanase skew information. Thus, we demonstrated that ISS methods can be used beyond metrological performance optimization in other important quantum information applications.  

\section{Acknowledgements} 

We thank I.~Apellaniz, R.~Augusiak, R.~Demkowicz-Dobrza\'nski, O.~G\"uhne,
P.~Horodecki, R.~Horodecki, and J.~Siewert for discussions.
We thank J.~Ko\l ody\'nski, K. Macieszczak, and R.~Demkowicz-Dobrza\'nski for discussing the details of \REF{Macieszczak2013Quantum_arxiv}.
We acknowledge the support of the  EU (COST Action CA23115, 
QuantERA MENTA, QuantERA QuSiED), the Spanish MCIU (Grant No.
PCI2022-132947), the Basque Government (Grant No.
 IT1470-22), and the National Research, Development and Innovation Office NKFIH
(Grant No. 2019-2.1.7-ERA-NET-2021-00036, Advanced Grant No. 152794).  
We acknowledge the support of the Grant No. PID2021-126273NB-I00 funded by
MCIN/AEI/10.13039/501100011033 and by "ERDF A way of making Europe".
We thank the
"Frontline" Research Excellence Programme of the NKFIH (Grant No. KKP133827).
We thank the Project No. TKP2021-NVA-04, which has been implemented with the support provided by the Ministry of Innovation and Technology of Hungary from the National Research, Development and Innovation Fund, financed under the TKP2021-NVA funding scheme. We thank the Quantum Information National Laboratory of Hungary. G.~T. thanks a  Bessel Research Award of the Humboldt Foundation.
We also acknowledge the support of the EU (QuantERA eDICT, CHIST-ERA MoDIC) and the
National Research, Development and Innovation Office NKFIH
(No.~2019-2.1.7-ERA-NET-2020-00003, No.~2023-1.2.1-ERA\_NET-2023-00009, No.~KH125096, and No.~K145927).

\appendix

\section{Special set of separable states similar to the set of local Hamiltonians}
\label{sec:specsep}

In this Appendix, we consider a set of separable states that are analogous to
local Hamiltonians, namely states of the form \be \label{eq:specSetDef} \varrho
= p \left( \varrho_1 \otimes \frac{\eins}{d_2} \right) + (1-p) \left(
\frac{\eins}{d_1} \otimes \varrho_2 \right)\,,
\ee
where $0\le p\le 1,$ $d_1$ and $d_2$ are the dimensions of the two Hilbert
spaces.

First, for completeness, we explicitly show that the set of states of the form
given in \EQ{eq:specSetDef} is convex. Let
\be
\varrho^{(k)} = p_k \left( \varrho_1^{(k)} \otimes \frac{\eins}{d_2} \right) +
(1-p_k) \left( \frac{\eins}{d_1} \otimes \varrho_2^{(k)}
\right)\label{eq:specsep}
\ee
for $k={\rm I,II}$, be two such states. Let us consider a density matrix being the mixture of $\varrho^{({\rm I})}$ and $\varrho^{({\rm II})}$
\be
\varrho = r \varrho^{({\rm I})} + (1-r)\varrho^{({\rm II})},
\ee
where $0\le r \le 1.$  We shall show, that there exist $\varrho_1$,
$\varrho_2$ and $p$ such that $\varrho$ can be written in the form given in \EQ{eq:specSetDef}. Let us
consider the unnormalized density matrices
\bea
\varrho_1' &=& r p_{\rm I} \varrho_1^{({\rm I})} + (1-r) p_{\rm II}
\varrho_1^{({\rm II})},\nonumber\\
\varrho_2' &=& r (1-p_{\rm I}) \varrho_2^{({\rm I})} + (1-r)(1-p_{\rm
II})\varrho_2^{({\rm II})},
\eea
Let us also define their traces as 
\be
t_1 = \trace(\varrho_1'),\quad t_2 = \trace(\varrho_2'),
\ee
Then, with $\varrho_1=\varrho_1'/t_1,$ $\varrho_2=\varrho_2'/t_2,$ $p=t_1$,
$\varrho$ can be formally written as in Eq.\ (\ref{eq:specSetDef}). We add that
both $\varrho_1$ and $\varrho_2$ are density matrices: combinations of positive
self-adjoint matrices with positive coefficients, thus positive, and of unit
trace. What remains is to check that $t_1 + t_2 =1$, which holds.

Let us now consider the recovery of $\varrho_1$ and $\varrho_2$ from a state of
the form (\ref{eq:specSetDef}). As a first step, the reduced density matrices
are
\begin{equation}\label{eq:redDM}
\begin{aligned}
\varrho^{\rm red}_1 &= \trace_2 \varrho = p \varrho_1 + (1-p)\frac{\eins}{d_1},\\
\varrho^{\rm red}_2 &= \trace_1 \varrho = (1-p) \varrho_2 + p \frac{\eins}{d_2}.
\end{aligned}
\end{equation}
It is possible to solve Eqs.\ (\ref{eq:redDM}) for $\varrho_1$ and $\varrho_2$
as
\begin{equation}\label{eq:specSetSol}
\begin{aligned}
\varrho_1 &= \frac{1}{p}\left( \varrho^{\rm red}_1 - \frac{1-p}{d_1}\eins\right),\\
\varrho_2 &= \frac{1}{1-p}\left( \varrho^{\rm red}_2 - \frac{p}{d_2}\eins\right).\\
\end{aligned}
\end{equation}
However, if $p$ is not known, the recovery is not unique. Eqs.\
(\ref{eq:specSetSol}) yield a solution for any $p$, for which $\varrho_1 \ge 0$
and $\varrho_2\ge 0$ hold, constraining $p$ as
\begin{equation}\label{eq:specSetBounds}
1-d_1 \sigma_{\rm min}(\varrho^{\rm red}_1) \le p \le d_2 \sigma_{\rm min}(\varrho^{\rm red}_2).
\end{equation}
Note, that the bounds (\ref{eq:specSetBounds}) also ensure that $0\le p \le 1$.
Based on \EQ{eq:specSetBounds}, a necessary condition for the existence of some
allowed $p$ is 
\be
1-d_1 \sigma_{\rm min}(\varrho^{\rm red}_1) \le d_2 \sigma_{\rm
min}(\varrho^{\rm red}_2),
\ee
otherwise the state cannot be written in the form given in \EQ{eq:specSetDef}.
It can also happen that \EQ{eq:specSetBounds} allows a single $p$ value. For
instance, if one of the reduced density matrices $\varrho^{\rm red}_1$ or
$\varrho^{\rm red}_2$ is not of full rank, only $p=1$ or $p=0$, respectively, is
possible.

Based on these, if a density matrix $\varrho$ is given only, and one needs to
know if it is of the form (\ref{eq:specSetDef}), one needs to choose a parameter
$p$ fulfilling the bounds (\ref{eq:specSetBounds}), calculate $\varrho_1$ and
$\varrho_2$, and afterwards verify, if formula (\ref{eq:specSetDef}) reproduces
the density matrix given. If it works for one allowed $p$ value, it will work
for the others as well, thus we need to try it only once. Thus, we can always decide
whether a state is of the form \EQ{eq:specSetDef}, while it is not possible to
decide efficiently whether a quantum state is separable or not \cite{Horodecki2009Quantum,Guhne2009Entanglement,Friis2019,Horodecki2021Quantum}.

Straightforward algebra shows that for states of the form \EQ{eq:specsep} and
for traceless $A$ and $B$ Hermitian operators the relation
\be
\ex{A \otimes B}=0 \label{eq:AB0}
\ee
holds. Note that \EQ{eq:AB0} does not mean that there are no correlations, since 
\be
\ex{A \otimes B}-\ex{A \otimes \openone}\ex{ \openone\otimes B}
\ee
 is not necessarily zero. Let us now consider a full basis of traceless
Hermitian matrices in the two subsystem 
\be
\{G_1^{(k)}\}_{k=1}^{s_1},\; \{G_2^{(k)}\}_{k=1}^{s_2},
\ee
with the number of elements given as 
\be
s_n=d_n^2-1
\ee
for $n=1,2.$
We assume that the basis matrices are pairwise orthogonal to each other
\be
{\rm Tr}(G_n^{(k)}G_n^{(l)})=2\delta_{kl},
\ee
where $n=1,2$ and $\delta_{kl}$ is the Kronecker symbol. Matrices $G_1^{(k)}$  and
 $G_2^{(l)}$ are the $SU(d)$ generators for $d=d_1,$ and $d=d_2,$ respectively.
 Then, any density matrix $\varrho$ can be expressed as a linear combination as 
\bea
\varrho&=&\frac{1}{d_1d_2}\openone\otimes\openone+\frac1 {2d_2}\sum_k \lambda_1^{(k)}
G_1^{(k)} \otimes \openone\nonumber\\
&+&\frac1 {2d_1}\sum_k \lambda_2^{(k)} \openone \otimes G_2^{(k)} + \frac1
{4}\sum_{k,l} K_{k,l} G_1^{(k)} \otimes G_2^{(l)},\nonumber\\
\eea
 for $1\le k \le s_1$ and $1\le l \le s_2.$ Here, $\lambda_1^{(k)},$
$\lambda_2^{(l)},$ and $K_{k,l}$ are real coefficients that form the elements of
the coherence vector of the system that describes the state of the system
equivalently to the density matrix \cite{Mahler1998Quantum}. Due to \EQ{eq:AB0},
most of the $d_1^2d_2^2-1$ elements of the coherence vector are zero since
\be
K_{k,l}=\ex{G_1^{(k)} \otimes G_2^{(l)}}=0\label{eq:K0}
\ee
holds for all $k,l.$ Thus, the number of nonzero elements is at most $s_1+s_2.$ The remaining elements of the coherence vector are given as 
\bea
\lambda_1^{(k)}&=&\ex{G_1^{(k)} \otimes \openone},\nonumber\\
\lambda_2^{(k)}&=&\ex{  \openone\otimes G_2^{(k)}}.
\eea
The following statement can straightforwardly be proven. If the state fulfills the condition in \EQ{eq:K0} then it can be written in the form given in \EQ{eq:specSetDef}. Thus, we can decide whether a state is of the form given in \EQ{eq:specSetDef} based on verifying that \EQ{eq:K0}  holds.

\section{See-saw algorithm with and without semidefinite programming}
 \label{app:opt}
 
 In this Appendix, we discuss how to implement a single step of the ISS
 method given in \SEC{sec:Bilinear}.
 
 Based on Observation 3, we need to obtain
 \be
\max_{\mathcal H \in \mathcal L_{c_1,c_2}} \trace[(Q \circ Q \circ \mathcal H_{\rm old}) \mathcal
H].
\label{eq:seesaw2}
\ee
Here $\mathcal  H_{\rm old}$ is the previous guess or the initial random guess. $\mathcal H$ is of the form
\EQ{eq:bipartiteH}, where $H_n$ fulfill \EQ{eq:cnHconst} for $n=1,2.$  This can
be solved by semidefinite programming.

\EQL{eq:seesaw2} can also be solved without semidefinite programming. The
quantity to be maximized in \EQ{eq:seesaw2} can be written as
  \be
 \FQ[\varrho,\mathcal H]=\trace(W_1 H_1) + \trace(W_2 H_2)\label{eq:max}
 \ee
where $W_n$ is given as 
\be
W_n=\trace_{\{1,2\}\backslash n} (Q \circ Q \circ \mathcal H_{\rm old}).
\ee
Let us write down the eigendecomposition of $W_n$ as
\be
W_n=U_n D_n U_n^\dagger,
\ee
where $U_n$ is unitary and $D_n$ is diagonal. We define the diagonal matrix
\be
(\tilde D_n)_{k,k} = c_k s[(D_n)_{k,k}],
\ee
where $s(x)=1$ if $x \ge 0,$ and $-1$ otherwise. With these the optimal
Hamiltonians are given by
\be
H_{n}^{({\rm opt})}=U_n \tilde D_n U_n^\dagger
\ee
for $n=1,2.$ From these, we can obtain the new $\mathcal H_{\rm old}$ and go to the next iteration step.
Note that a similar scheme was presented in another context in
\REF{Toth2020Activating}.

\section{Alternative method to optimize quadratic functions}
\label{app:alt}

In this Appendix, we present an alternative of the method given in
\SEC{sec:Bilinear}. We need the optimization over a constrained and an
unconstrained quantity, while in the method of \SEC{sec:Bilinear}, we need to
optimize over two constrained quantities.

Let us try to maximize the convex function 
\be
f(x)=4q^2x^2\label{eq:gx}
\ee
on a convex set. We could check the extreme point of the set, however, we look
for a method that can easily be generalized to a larger number of variables.

\begin{figure}[t!]
\includegraphics[width=\columnwidth]{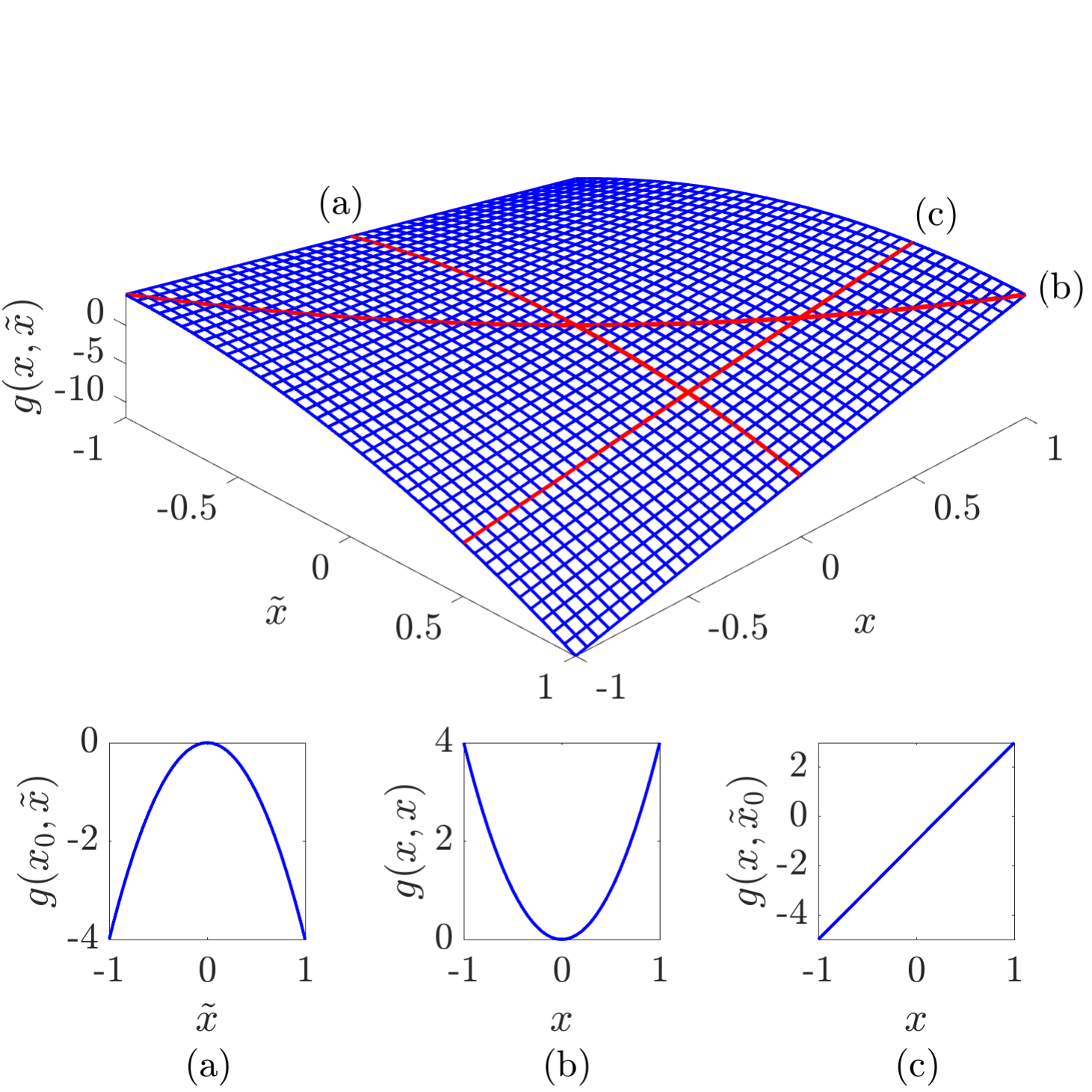}
\caption{Plot of $g(x,\tilde x)$ as defined in \EQ{eq:gxy} for $q=1.$ In the insets, curves for $x_0=0$ and $\tilde x_0=0.5$ are shown on the sides, while the middle inset is for $\tilde x=x.$ Red lines in the main figure correspond to the three insets. By inspection we can see that $g(x,\tilde x)$ is not concave.} \label{fig:convfig}
\end{figure}

We introduce another function with an auxiliary variable $\tilde x$ that is easier to
maximize. The linear function tangent to $f(x)$ at the point $x=\tilde x$ is given as
\be
g(x,\tilde x)=t(\tilde x)+\frac{df}{dx}\bigg\vert_{x=\tilde x}\times x=-4q^2\tilde x^2+8q^2\tilde x x,\label{eq:gxy}
\ee
where the quantity
\be
t(\tilde x)=f(\tilde x)-\frac{df}{dx}\bigg\vert_{x=\tilde x} \tilde x
\ee
 is the value of the linear function at $x=0,$  which equals $-1$ times the Legendre transform of $f(x)$ \cite{Rockafellar2015Convex,Boyd2004Convex,Zia2009Making,Guhne2007Estimating,Apellaniz2017Optimal}.
We can maximize $f(x)$ given in \EQ{eq:gx} as 
\be
\max_{x\in \mathcal X} f(x) =\max_{x\in \mathcal X} \max_{\tilde x} g(x,\tilde x),
\ee
where $\mathcal X$ is a convex set of real numbers, and $y$ is real. 
$g(x,\tilde x)$ is plotted in \FIG{fig:convfig} for $q=1.$

Then the maximization can be carried out as follows. 

(i) Choose a random $x$.

(ii) Maximize $g(x,\tilde x)$  over $\tilde x$ for a fixed $x.$ We have to use the condition
for maximum
\be
\frac{\partial g(x,\tilde x)}{\partial \tilde x}=8q^2(x-\tilde x)=0.
\ee
Hence, we have to set the auxiliary variable $\tilde x$ to
\be
\tilde x=x.\label{eq:y12qx}
\ee

(iii) Maximize $g(x,\tilde x)$  over $x$ for a fixed $\tilde x.$ The maximum will be taken at
an extreme point of $\mathcal X.$ After completing step (iii), go to step (ii).

The function $g(x,\tilde x)$ is linear in $x,$ concave in $\tilde x.$  The
Hessian matrix of the second order derivatives is 
 \be
D= \left(\begin{array}{ll}0 & \phantom{-}8q^2 \\8q^2 & -8q^2\end{array}\right).
 \ee
 The eigenvalues of the Hessian are 
 \be
  \lambda_{\pm}=(\pm 4\sqrt{5} - 4)q^2.
 \ee
 One of the eigenvalues is positive for any $q>0.$ Thus, $g(x,\tilde x)$ is not concave in
$(x,\tilde x)$ since the Hessian has positive eigenvalues. Thus, it is not guaranteed
 that we will find a unique maximum of a concave function.

Let us now generalize these ideas to several variables. Let us assume that
$\mathcal H \in \mathcal L_{c_1,c_2}$ and $\varrho$ is a density matrix. Then,
we can write that 
\be
\max_{\mathcal H \in  \mathcal L_{c_1,c_2}} \FQ[\varrho,\mathcal H]
=\max_{\mathcal H \in  \mathcal L_{c_1,c_2}} \max_{\tilde{\mathcal H}} G(\mathcal H,\tilde{\mathcal H}),
\ee
where $G(\mathcal H,\tilde {\mathcal H})$ is defined as
\bea
&&G(\mathcal H,\tilde {\mathcal H}) \nonumber\\
&&\quad\quad= \sum_{kl} Q_{kl}^2 \left[\tilde {\mathcal H}^{\rm r}_{kl}(2{\mathcal H}_{kl}^{\rm r}-\tilde {\mathcal H}_{kl}^{\rm
r})+\tilde {\mathcal H}^{\rm i}_{kl}(2{\mathcal H}_{kl}^{\rm i}-\tilde {\mathcal H}_{kl}^{\rm
i})\right].\nonumber\\
\label{eq:GHY}
\eea
Note the difference compared to the bilinear form in \EQ{eq:seesaw}: The optimization over $\tilde H_{kl}$ is unconstrained.

\section{Optimization over the state rather than the Hamiltonian}
\label{app:state}

In this Appendix,
we will analyze the method optimizing the quantum Fisher information over the quantum state,
rather than over the Hamiltonian, described in 
\REF{Macieszczak2013Quantum_arxiv,Macieszczak2014Bayesian,Chabuda2020Tensor}.
We would like to examine whether it finds the global optimum.
It turns out that the see-saw used for optimization is similar to that of \APP{app:alt}.
Thus, we will apply an analysis similar to the one presented in \APP{app:alt}.

Let us review the method briefly. 
We need to maximize the error propagation formula
\begin{eqnarray}
\max_{{\varrho}} \FQ[\varrho,\mathcal H]&=&\max_{{\varrho}}\max_M
1/{\va{\theta}_M}\nonumber\\
&=&\max_{\varrho} \max_{M} {\ex{i[{\mathcal H},M]}^2}/{\va{M} }\nonumber\\
&=&\max_{\varrho} \max_{M} {\ex{i[{\mathcal H},M]}^2}/{\ex{M^2} },\label{eq:fmin0}
\end{eqnarray}
where the last equality stands, since $M$ operators for which $\ex{M}_\varrho=0$
maximize both $\va{M}$ and $\ex{M^2},$ and for such operators $\va{M}=\ex{M^2}.$
Then, maximizing a fraction over $M$ such that $M$ appears both in the numerator and the denominator is a difficult task. Thus, a variational approach has been used to obtain the optimum as \cite{Macieszczak2013Quantum_arxiv}
\begin{eqnarray}
\max_{{\varrho}} \FQ[\varrho,\mathcal H]&=&\max_{\varrho} \max_M \max_\alpha \{ -\alpha^2 \ex{M^2} + 2 \alpha
\ex{i[{\mathcal H},M]}  \}.\nonumber\\\label{eq:fmin}
\end{eqnarray}
For the maximum \cite{Macieszczak2013Quantum_arxiv}
\bea
&&\frac{\partial}{\partial\alpha} \left( -\alpha^2 \ex{M^2} + 2 \alpha \ex{i[{\mathcal H},M]} \right)\nonumber\\
&&\quad\quad\quad=
 -2\alpha \ex{M^2} + 2 \ex{i[{\mathcal H},M]} =0
\eea
holds. Then, for the optimum $\alpha=\ex{i[{\mathcal H},M]} / \ex{M^2}$ and substituting this into \EQ{eq:fmin}, we obtain the right-hand side of \EQ{eq:fmin0}. If we carry out the optimization over $\alpha$ and $M$ then we find that the optimum is taken when $M$ equals the symmetric logarithmic derivative and $\alpha=1.$

In order to decrease the number of parameters we need to optimize, the optimization in \EQ{eq:fmin} can be rewritten as 
\be
\max_{{\varrho}} \FQ[\varrho,\mathcal H]=\max_{\varrho} \max_{M'} \{ -\ex{(M')^2} + 2 \ex{i[{\mathcal H},M']}\},\label{eq:fmin2}
\ee
where $M'$ takes the role of $\alpha M.$ Note that the function in \EQ{eq:fmin2} is concave in $M'$
and linear in $\varrho.$ 

The expression in \EQ{eq:fmin2} can be used to define a
two-step see-saw algorithm that will find better and better quantum states. 
First we choose a random state. Then, maximize the expression over $M'.$ One can show that the maximum is taken by setting $M'$ to be the symmetric logarithmic derivative given in \EQ{eq:SLD} \cite{Macieszczak2013Quantum_arxiv}. Then, we optimize over $\varrho.$ Then, we go back to the step maximizing over $M',$ and so on. The
optimization seems to work very well, converging almost always in practice. 

\begin{figure}[t!]
\includegraphics[width=\columnwidth]{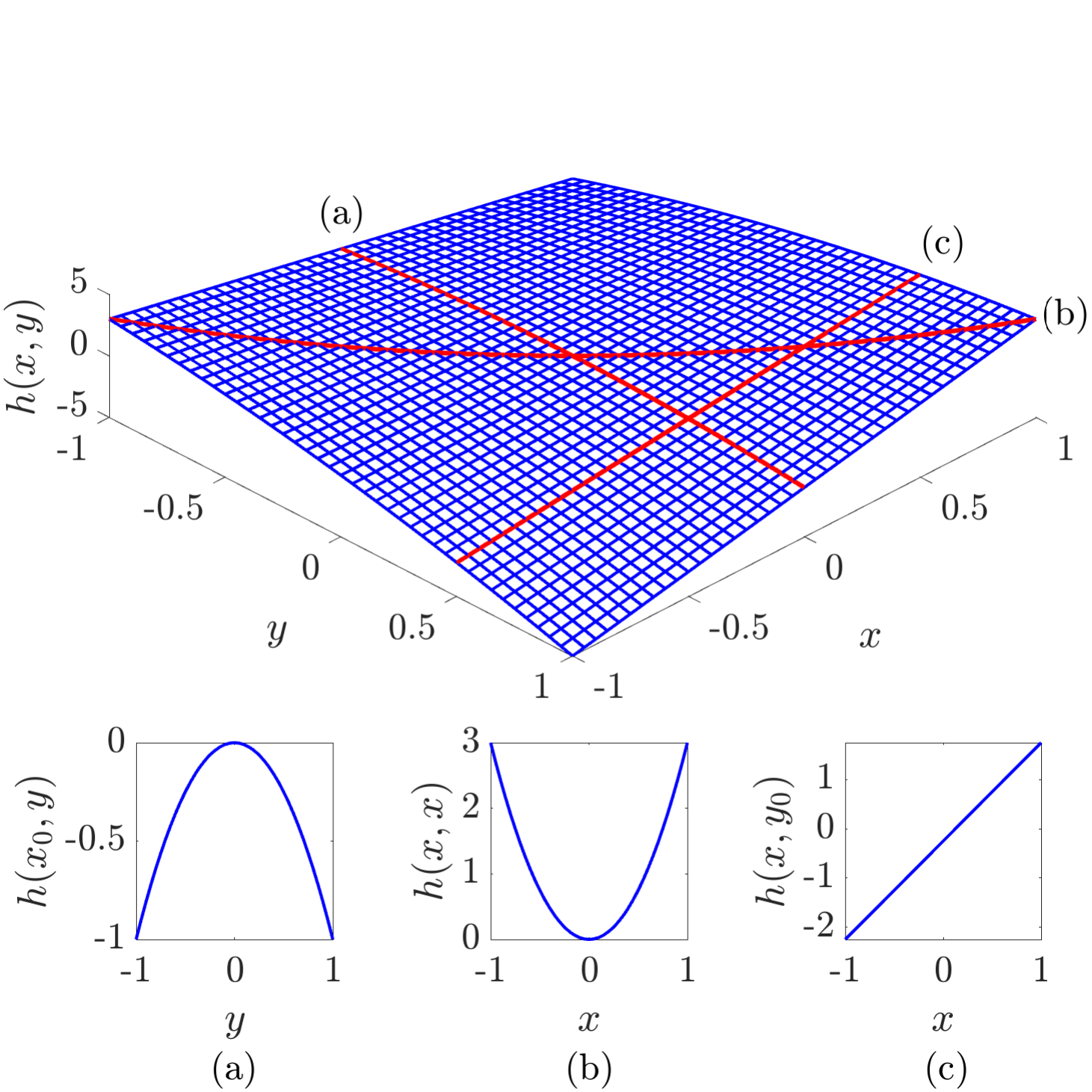}
\caption{Plot of $h(x,y)$ as defined in \EQ{eq:expr2}. In the insets, curves for $x_0=0$ and $y_0=0.5$ are shown on the sides, while the middle inset is for $y=x.$ Red lines in the main figure correspond to the three insets. By inspection, we can see that $h(x,y)$ is not concave. } \label{fig:convfig_opt_over_rho}
\end{figure}

After reviewing the powerful method in \REFS{Macieszczak2013Quantum_arxiv,Macieszczak2014Bayesian,Chabuda2020Tensor}, the
question arises: do we always find the global maximum? For that we need that the expression maximized in \EQ{eq:fmin2}
is strictly concave in $(\varrho,M').$ 
Let us see a concrete example. Let us consider a single qubit with a density
 matrix 
  \be
 \varrho=\frac1 2(\openone + \sigma_x x),\label{eq:rho1d}
 \ee
 where the allowed values for the real parameter $x$ are given by
 \be
 x_{\min} \le x \le x_{\max},\label{eq:interval}
 \ee 
 and $[x_{\min},x_{\max}]$ is a subset of the $[-1,1]$ interval.
 Clearly,
\be
\ex{\sigma_x}_\varrho=x.
\ee
We will carry out an optimization of the quantum Fisher information over the convex set of states defined above. Let us now consider the Hamiltonian
 \be
\mathcal H=\sigma_z.
 \ee
Then, we set $M'$ to the symmetric logarithmic derivative defined in \EQ{eq:SLD} to
 \be
 M'=\sigma_y y,
 \ee
 where $y$ is the auxiliary, real parameter. The expression maximized in \EQ{eq:fmin2} equals
 \be
 h(x,y)=-y^2+4xy,\label{eq:expr2}
 \ee
 which is maximized for a given $x$ if $y=2x.$ Then, we can eliminate $y$ and obtain 
  \be
\FQ[\varrho,\mathcal H]=4x^2.\label{eq:expr2_e}
 \ee
 The Hessian matrix of the second order derivatives of the expression in \EQ{eq:expr2} is 
 \be
D= \left(\begin{array}{ll}0 & \phantom{-}4 \\4 & -2\end{array}\right).
 \ee
 Since one of the two eigenvalues of the Hessian is negative, the other is positive, the expression in \EQ{eq:expr2}  is not concave in $x$ and $y.$ In fact, \EQ{eq:expr2} is concave in $y$ and linear in $x,$ but not concave in $(x,y).$ The function $h(x,y)$ given in \EQ{eq:expr2} is plotted in \FIG{fig:convfig_opt_over_rho}, which makes it possible to understand the convexity properties of  \EQ{eq:expr2}. Thus, the expression maximized in \EQ{eq:fmin2} is in general not concave in $(\varrho,M').$

Following the ideas of \APP{app:alt} and applying the Legendre transform \cite{Rockafellar2015Convex,Boyd2004Convex,Zia2009Making,Guhne2007Estimating,Apellaniz2017Optimal}, the equation of the tangent to the function given in \EQ{eq:expr2_e} is
\be
h(x,\tilde x)=-4\tilde x^2+8\tilde x x.\label{eq:gprime}
\ee
One can use an optimization over $\tilde x$ to obtain  
\be
\FQ[\varrho,\mathcal H] = \max_{\tilde x}h(x,\tilde x)=4x^2.
\ee
Now, by substituting $\tilde x =y/2$ into \EQ{eq:gprime} we arrive at \EQ{eq:expr2}. Following the ideas of \APP{app:alt}, we can also see that \EQ{eq:fmin2} is also based on linearizing $\FQ[\varrho,\mathcal H]$ at a given $\varrho,$ which leads to a tangent plane, and then maximizing the quantum Fisher information over that tangent plane.

Let us consider a maximization of \EQ{eq:expr2} without further constraints such that $x_{\min}=-1$ and $x_{\max}=+1.$ We start from a random $\varrho,$ which in our case means to start with a random $x.$ Let us consider the case that $x>0.$ Then, we will have $y=2x>0.$ Then, when optimizing over $x$ in the next step, we get $x=+1.$ Let us now consider the case that $x<0.$ Then, we will have $y=2x<0.$ Then, when optimizing over $x,$ we get $x=-1.$ Thus, we will find the correct maximum, independently from the initial state. The reason is that the Fisher information is invariant under the transformation $x\rightarrow -x.$ 

{\bf Observation 6.} In the single-qubit case, the see-saw always reaches the global maximum for any initial state, if we optimize over the full set of physical quantum states.
 
{\it Proof.} Let us consider a single qubit in a general state
  \be
 \varrho=\frac1 2(\openone + r_x\sigma_x  + r_y \sigma_y  + r_z \sigma_z),
 \ee
 and a general single-qubit $M'$ as
  \be
 M'=m_x\sigma_x + m_y\sigma_y+m_z\sigma_z .
 \ee
 With these, the quantum Fisher information can be given as 
 \be
\FQ[\varrho,\mathcal H]=\max_{m_x,m_y,m_z}4(m_yr_x  - m_x r_y) - m_x^2 - m_y^2 - m_z^2.
 \ee
 The maximum is taken at $m_x=-2r_y, m_y=2r_x,$ and $m_z=0.$
 After eliminating $m_k$ we obtain
 \be
\FQ[\varrho,\mathcal H]= 4(r_x^2 + r_y^2),\label{eq:FQrxry}
 \ee
which takes the global maximum for all points for which $r_x^2+r_y^2=1$ and $r_z=0.$ 
Based on these we see that starting from any random initial state, we always reach the global maximum, except for the case when $r_x=r_y=0,$ when we obtain
$m_x=m_y=m_z=0.$ However, such states form a zero measure set, thus small additive noise with a random quantum state leads out of the set,
and then the algorithm finds the global maximum. $\qed$

In the more general case of larger systems, the quantum Fisher information is not a quadratic function of the density matrix elements anymore \cite{Safranek2018Simple,Rath2021Quantum}. However, even in this case,  if $\varrho_{\rm opt}$ is maximizing the quantum Fisher information $\FQ[\varrho,\mathcal H],$ then 
\be
\varrho_{\rm opt}'=\exp(-iK)\varrho_{\rm opt}\exp(+iK)
\ee
 also maximizes it, where $[\mathcal H,K]=0.$ Thus, even though the final state of the iteration depends on the initial state, the algorithm finds the global maximum. We tested the optimization problem for physical quantum states of various sizes ($2\times2$, $3\times3$, ..., $12\times12$). In this case, looking for the maximum of $\varrho$ in \EQ{eq:fmin2} means identifying the eigenstate with the maximal eigenvalue of the operator \cite{Macieszczak2013Quantum_arxiv}
\be
G(M')=-(M')^2+2i[\mathcal H,M'].\label{eq:GMprime}
\ee
For a pure state $\ket{\Psi},$ the symmetric logarithmic derivative for a given Hamiltonian is given in \EQ{eq:SLDpure}. Substituting it into $M'$, we obtain
\be
G_{\Psi}=4(\{\varrho,\mathcal H\}\ex{\mathcal H}_\varrho-\varrho\ex{\mathcal H^2}_\varrho +\{\varrho,\mathcal H^2\}-3\mathcal H\varrho \mathcal H),\label{eq:Gpsi2}
\ee
where $\varrho=\ketbra{\Psi}.$ 

If the initial state $\ket{\Psi}$ is an eigenvector of $\mathcal H,$ then $G_{\Psi}=0,$  i.~e., a matrix with zero elements. Thus, the next state of the iteration obtained as the eigenvector of $G_{\Psi}$ with the largest eigenvalue will be a random vector.

If the initial state $\ket{\Psi}$ is
\be
\ket{\Psi}=\frac1 {\sqrt{2}}  (\ket{h_{k}}+e^{-i\phi}\ket{h_{l}}),\label{eq:maxim2}
\ee
then we have
\bea
G_{\Psi}&&=(h_k-h_l)^2\bigg[2(e^{+i\phi} \ket{h_k}\bra{h_l}+e^{-i\phi} \ket{h_l}\bra{h_k})\nonumber\\
&&-\ket{h_k}\bra{h_k}-\ket{h_l}\bra{h_l}\bigg].\label{eq:maxim222}
\eea
Here,  $h_k$ and $h_l$ denote the eigenvalues of $\mathcal H,$ 
while $\ket{h_k}$ and $\ket{h_l}$ are corresponding eigenvectors. The eigenvector of \EQ{eq:maxim222} with the largest eigenvalue is the state given in \EQ{eq:maxim2}. 

Let us see now the steady states of the see-saw. For such states,
\be
G_{\Psi}\ket{\Psi}=4(\mathcal H-\ex{\mathcal H})^2\ket{\Psi}=\lambda\ket{\Psi}\label{eq:Glambda}
\ee
holds, where $\lambda$ is the maximal eigenvalue of $G_{\Psi}$ given in \EQ{eq:Gpsi2}.
Thus, every steady state must fulfill the condition given in \EQ{eq:Glambda} for some $\lambda.$
Let us consider first the case that $\mathcal H$ has a non-degenerate spectrum.
Let us look for the eigenvectors of the operator
\be
4(\mathcal H-c)^2,
\ee
where $c$ is a constant. 
For $c=(h_k+h_l)/2,$ the solution is of the form
$\alpha\ket{h_k}+\beta\ket{h_l},$ where $\alpha$ and $\beta$ are constants. For these states,
$\ex{\mathcal H}=|\alpha|^2h_k+|\beta|^2h_l.$
For all $c,$ all eigenstates $\ket{h_k}$ of $\mathcal H$ are solutions.  
The solutions for which $\ex{\mathcal H}=c$ are states of the type $\ket{h_k},$ and states given in \EQ{eq:maxim2}.

The case of $\ket{h_k}$ we discussed in relation to \EQ{eq:Gpsi2}.
We find that states of the type given in \EQ{eq:maxim2} are stable points of the iteration. However, based on numerics, we find that a small perturbation outside of the subspace $\{\ket{h_k},\ket{h_l}\}$
will move the iteration away from the state, unless we are in the state
\be
\ket{\Psi_{\rm opt}}=\frac1 {\sqrt{2}}  (\ket{h_{\min}}+e^{-i\phi}\ket{h_{\max}}),\label{eq:maxim}
\ee
which is stable under perturbations.
Here, $\phi$ is a real number and $\ket{h_{\min}}$ and $\ket{h_{\max}}$ are the eigenvectors with the minimum and maximum eigenvalues of $\mathcal H.$ 
If $\mathcal H$ has a degenerate maximal or minimal eigenvalue, then the steady states can also be obtained straightforwardly.

The steady states maximize
\be
\ex{G_{\Psi}}_{\Psi}=4\va{\mathcal H}_{\Psi}.\label{eq:Gpsi}
\ee
Thus, the mechanism of the see-saw resembles the power iteration discussed in \SEC{eq:poweriteration}, however, now we maximize the variance of an operator. 
It has been shown that the iteration leads to states with a variance $\va{\mathcal H}$ that does not decrease \cite{Macieszczak2013Quantum_arxiv}.
Together with our argument that the steady states all maximize the variance, we can see that the iteration always finds the global maximum.
We verified this claim based on extensive numerical tests \footnote{In principle, it is possible to find an initial state for which $M',$ calculated by \EQ{eq:SLD}, equals the zero matrix. This is the case if $\mathcal H$ is not in the range of the initial density matrix.
However, small random added noise can easily help, and the iteration will lead to the global maximum.}.

Next, we can consider the set of quantum states obtained via a physical map
$\Lambda(\varrho)$ \cite{Macieszczak2013Quantum_arxiv}. The first example is the states from the globally depolarizing map.
We have to look for the eigenvector with the maximum eigenvalue
in the following eigenequation \cite{Macieszczak2013Quantum_arxiv}
\be
\Lambda^\dagger(G_{\varrho_n})\ket{\Psi}=\lambda\ket{\Psi},\label{eq:Gpsi_lambda}
\ee
where for computing $G_{\varrho_n}$, we used the symmetric logarithmic derivative of the noisy input state $\varrho_n$ given as  
\be
\varrho_n=\Lambda(\ketbra{\Psi})=p\ketbra{\Psi}+(1-p)\frac{\openone}{d}.
\ee
The symmetric logarithmic derivative for $\Lambda(\ketbra{\Psi})$ is given as
$
qM',
$
where $M'$ is the symmetric logarithmic derivative  for the noiseless pure state, $q=p/[p+2(1-p)/d],$ and $q=1$ corresponds to no noise.
Hence, we obtain
\be
G_{\varrho_n}(M')=-q^2(M')^2+q2i[\mathcal H,M'],
\ee
c.~f. \EQ{eq:GMprime}. The adjoint map is
\be
\Lambda^\dagger(A)=pA+(1-p)\frac{\openone}{d}{\rm Tr}(A).
\ee
Then, we arrive at
\begin{align}
&\Lambda^\dagger(G_{\varrho_n})=-pq^2(M')^2+pq2i[\mathcal H,M']+(1-p)\frac{\openone}{d}{\rm Tr}(G_{\varrho_n}).
\end{align}
Hence, substituting $M'$ with the symmetric logarithmic derivative given for pure states in \EQ{eq:SLDpure}, for $\Lambda^\dagger(G_{\varrho_n})\ket{\Psi}$ we obtain
\begin{align}
\Lambda^\dagger(G_{\varrho_n})\ket{\Psi}&=\bigg[4pq(1-q)\va{\mathcal H}_\varrho\nonumber\\
& +4pq(\mathcal H-\ex{\mathcal H}_\varrho)^2\nonumber\\
&+\frac{1-p}{d}{\rm Tr(G_{\varrho_n})}\bigg]\ket{\Psi}.
\end{align}
The part that contributes to finding the eigenvector is 
\be
4pq(\mathcal H-\ex{\mathcal H}_\varrho)^2,
\ee
c.~f. \EQ{eq:Glambda}.
After a derivation similar to the previous one, we find again that the algorithm is maximizing $\va{\mathcal H}_{\varrho}$ and the same states maximize it as before.
Thus, based on the same arguments, the algorithm always finds the global maximum.

Next, we present a case when the algorithm might not find the global optimum, while admittedly the example is not a very practical one, and is constructed to test the method.
Let us consider a concrete iteration on a subset of physical quantum states of the type given in \EQ{eq:rho1d}, and assume $-1\le x_{\min}<0<x_{\max}\le 1$, which can be incorporated in the optimization as the condition
\be
x_{\min}\le \ex{\sigma_x}_{\varrho}\le x_{\max}. \label{eq:cond}
\ee

The inequality in \EQ{eq:cond} can be added as conditions to an optimization using semidefinite programming. Alternatively, we can also define a set of quantum states
fulfilling \EQ{eq:cond} via the map
\begin{align}
\Lambda_{1D}(\varrho)&=p_P (P_{x,-1}\varrho P_{x,-1}+P_{x,+1}\varrho P_{x,+1})\nonumber\\
&+p_{x,-1}P_{x,-1}+p_{x,+1}P_{x,+1}
\end{align}
from the set of quantum states, where $P_{x,+1}$ and $P_{x,-1}$ are projectors to the eigenstates of $\sigma_x.$  The minimal and maximal value for $\ex{\sigma_x}_{\varrho},$ respectively, are
\begin{align}
x_{\min}&=-p_P+p_{x,+1}-p_{x,-1},\nonumber\\
x_{\max}&=p_P+p_{x,+1}-p_{x,-1}.
\end{align}
Based on these, we can obtain the probabilities given with $x_{\min}$ and $x_{\max}$ as
$p_{x,-1}=(1-x_{ \max})/2,$ $p_{x,+1}=(1+x_{\min})/2,$ and $p_P =1-p_{x,-1}-p_{x,+1}.$

We start from a random $\varrho,$ that is, we start with a random $x.$ Let us consider the case that $x>0.$ Then, we will have $y=2x>0.$ Then, when optimizing over $x$ in the next step, we get $x=x_{\max}.$ Let us now consider the case that $x<0.$ Then, we will have $y=2x<0.$ Then, when optimizing over $x,$ we get $x=x_{\min}.$ Thus, depending on the sign of the random initial $x,$ we will arrive at  $x=x_{\min}$ or at  $x=x_{\max},$ which is not necessarily the global optimum.
  
Considering a single qubit in a general state, conditions such as
\bea
x_{\min}\le \ex{\sigma_x}_{\varrho}\le x_{\max},\nonumber\\
y_{\min}\le \ex{\sigma_y}_{\varrho}\le y_{\max},\label{eq:2D}
\eea
where $y_{\min}$ and $y_{\max}$ are real numbers giving the bounds of the allowed region for $\ex{\sigma_y}_{\varrho}$, can result in reaching different local optima starting from different initial states, even if small random perturbations are added. The proof is similar to the proof of the one-dimensional case discussed before. For example if $x_{\min}=y_{\min}=-1,$  $x_{\max}=y_{\max}=0.15,$ and the initial state is 
\be
\varrho'=\frac1 2(\openone + 0.05 \sigma_x  + 0.05 \sigma_y -0.02  \sigma_z ),\label{eq:intial}
\ee
then the optimization leads to $r_x=x_{\max},$ $r_y=y_{\max},$ and, based on \EQ{eq:FQrxry},  we have $\FQ[\varrho,\mathcal H]=4(x_{\max}^2+y_{\max}^2)=0.18.$ Thus,  it does not reach the maximal value $4.00.$

The inequalities in \EQ{eq:2D} can be added as conditions to an optimization using semidefinite programming. Alternatively, we can also define a set of quantum states via the physical map, e.~g.,
\begin{align}
\Lambda_{\rm qubit}(\varrho)&=p_I \varrho+\frac1 2 (1-p_I)(P_{x,-1}+P_{y,-1}) ,
\end{align}
where $P_{x,-1}$ is the projector to the eigenstate of $\sigma_x$ with eigenvalue $-1,$ and
 $P_{y,-1}$ is the projector to the eigenstate of $\sigma_y$ with eigenvalue $-1.$
For states coming from this map, 
\bea
 -\frac{p_I}2-\frac{1}{2}&\le& \ex{\sigma_x}_{\varrho}\le \frac{3p_I}2-\frac{1}{2}, \nonumber\\
 -\frac{p_I}2-\frac{1}{2}&\le& \ex{\sigma_y}_{\varrho}\le \frac{3p_I}2-\frac{1}{2} \label{eq:2Db}
\eea
holds.  If the optimization is carried out over the quantum states coming from the map $\Lambda_{\rm qubit}(\varrho)$ then the 
maximum found will depend on the initial state.

More practical, relevant subsets of physical quantum states are PPT states in a bipartite system. Metrology with such states have been studied using semidefinite programming \cite{Toth2018Quantum}. A family of bipartite PPT entangled states have been found numerically maximizing the quantum Fisher information \cite{Toth2018Quantum}. Here, semidefinite programming was used to maximize over the set of PPT quantum states. Later, PPT states having the same quantum Fisher information have been found analytically, which is a strong indication that the method found the global maximum \cite{Pal2021Bound}.

In multiparticle systems, one can consider states that are PPT with respect to all bipartitions. Such four-qubit states have been studied \cite{Toth2018Quantum}. Such states form a very complicated set, and thus they are good candidates to test the method with a difficult task. We find that the maximum found in such systems might depend on the initial state. However, after starting from a couple of random initial values the iteration finds the global maximum. 

One might think that the semidefinite solvers can cause the dependence on the initial states. It is possible to consider a subset of physical quantum states such that semidefinite solvers are not needed. Such a set is the set of states with a symmetric extension of a given order \cite{Doherty2004Complete}. We considered the optimization over 2-symmetric extendible states and we also found a dependence on the initial state \cite{Navascues2011Activation}.

The optimization over PPT states can be reformulated with PPT-inducing channels acting on the set of quantum states \cite{Filippov2014PPT-Inducing}, which are similar to entanglement breaking maps  \cite{Horodecki2003EntanglementBreaking}.
The set of the 2-symmetric extendible states can be obtained from the set of all quantum states with a physical map.

In summary, in most relevant multi-qudit problems, the method seems to converge extremely well, which is very likely due to the fact that the global maximum is taken by many states. It is also important that several auxiliary variables are added. When considering states passing through a physical map, the symmetry properties of the map determine how well the method converges.
It would be interesting to prove that the approach always converges for several classes of relevant problems.

\section{An example for using the moment method to upper bound the quantum Fisher information}
\label{app:FQ_MOMENT_EXAMPLE}

In this Appendix, we exemplify the use of the moment method discussed in Sec.~\ref{sec:Quadprog} to upper bound the quantum Fisher information for a fixed $d\times d$ state $\varrho$. We will focus on $d=2$, however, the method generalizes straightforwardly to any $d>2$. 

Let us first recap the optimization problem. Fix a $2\times 2$ state $\varrho$ with the eigendecomposition  as in \EQ{eigdecomp}.
According to Eq.~\eqref{eq:QFI}, our task is to maximize 
\be
\FQ[\varrho,\mathcal H]= \sum_{k,l}Q_{kl}^2 \vert \mathcal H_{kl} \vert^2
\label{eq:QFIapp}
\ee
over bipartite local Hamiltonians of the form given in \EQ{eq:bipartiteH},
where we impose the condition \be{\rm Tr}(H_1^2) + {\rm Tr}(H_2^2)=4.\label{eq:TrTr4}\ee  The matrix elements of $\mathcal H$ in the eigenbasis of $\varrho$ are given by 
$\mathcal H_{kl}=\langle k|\mathcal H|l\rangle$. Let us denote by $\FQ^{(\text{max})}$ the maximum value in Eq.~\eqref{eq:QFIapp} over local Hamiltonians. Note that the above condition is a relaxation of the stricter one 
given in \EQ{eq:cnHconst2} for
 $n=1,2$ with $c_1^2 + c_2^2=2$. Therefore, $\FQ^{(\text{max})}$ in general defines only an upper bound to the true maximum of the quantum Fisher information. 
The constant coefficients $Q_{kl}$ depend only on the eigenvalues of the probe state $\varrho$ and are given by \EQ{eq:qqq}
whenever the denominator is nonzero. 

Now we apply the following steps to bring the problem to an SDP.

  (i) Build an orthonormal basis of Hermitian $d\times d$ matrices. For $d=2$, we choose the four normalized Pauli matrices $\hat{\sigma}_i = \tfrac{1}{\sqrt{2}}\sigma_i$, where $\sigma_0$ is the $2\times 2$ identity, and the rest are $\sigma_x$, $\sigma_y$, and $\sigma_z$. 
 
  (ii) Represent the local Hamiltonian operators as 
  \begin{equation}\label{eq:H12}
  \begin{aligned}
  H_1 &= \sum_{i=1}^{4} v_i \,\hat{\sigma}_{i-1},\\
  H_2 &= \sum_{i=1}^{4} v_{i+4} \,\hat{\sigma}_{i-1}
  \end{aligned}
  \end{equation}
  with real coefficients $v_i$, $i=1,\ldots,8$, forming the vector $\ket{v} \in \mathbb{R}^8$. With this choice we have 
  \be
  \braket{v}{v}=\mathrm{Tr}(H_1^2)+\mathrm{Tr}(H_2^2)=4,
  \ee 
 that is, we recover the condition in \EQ{eq:TrTr4}.
 
  (iii) For each $(k,l)$, define the matrices $S_{kl}=\mathrm{Tr}_B\ket{k}\bra{l}$ and $T_{kl}=\mathrm{Tr}_A\ket{k}\bra{l}$, and from them construct the complex vector $\ket{z_{kl}} \in \mathbb{C}^8$ compactly as
\begin{equation}
\begin{aligned}
\ket{z_{kl}} 
=\big[&\mathrm{Tr}\big(S_{kl}\hat{\sigma}_0\big),\mathrm{Tr}\big(S_{kl}\hat{\sigma}_x\big),\mathrm{Tr}\big(S_{kl}\hat{\sigma}_y\big),\mathrm{Tr}\big(S_{kl}\hat{\sigma}_z\big),   \\
&\mathrm{Tr}\big(T_{kl}\hat{\sigma}_0\big),\mathrm{Tr}\big(T_{kl}\hat{\sigma}_x\big),\mathrm{Tr}\big(T_{kl}\hat{\sigma}_y\big),\mathrm{Tr}\big(T_{kl}\hat{\sigma}_z\big)\big]^T.
  \end{aligned}
\end{equation}

(iv) In terms of $\ket{z_{kl}}$ we can write the complex matrix element as
\be
\mathcal H_{kl}=\langle k|\mathcal H|l\rangle=\mathrm{Tr}(S_{kl}H_1)+\mathrm{Tr}(T_{kl}H_2)=\braket{z_{kl}}{v}.
\ee

(v) Hence the quantum Fisher information is
\bea\FQ[\varrho,\mathcal H] &=& \sum_{k,l}Q_{kl}^2 \vert \mathcal H_{kl} \vert^2\nonumber\\
&=&\sum_{k,l}Q_{kl}^2\braket{v}{z_{kl}}\braket{z_{kl}}{v}=\braket{v}{R|v},\eea where  the operator
\begin{equation}
    R=\Re{\left(\sum_{kl}\ket{z_{kl}}\bra{z_{kl}}\right)} 
    \label{eq:R}
\end{equation}
is a real symmetric $8\times 8$ matrix, and we used that $\ket{v}$ is real valued.

This leads us to the optimization problem
\begin{equation}
\begin{aligned}
\FQ^{(\text{max})}:=\text{max}_{v\in\mathbb{R}^8}\quad & \langle v| R |v\rangle\\
\text{subject to}\quad & \langle v | v\rangle = 4.
\end{aligned}
\label{eq:maxFQv}
\end{equation}
Thus, we arrived at an optimization task of the form given in Eq.~\eqref{eq:blinear}. However, we can reformulate this problem as an SDP 
\begin{equation}
\begin{aligned}
\FQ^{(\text{max})}:=\text{max}_{\,X\in\mathbb S^8}\quad & \operatorname{Tr}(R X)\\[4pt]
\text{subject to}\quad & X \ge 0,\\[4pt]
& \operatorname{Tr}(X)=4,\\[4pt]
& \operatorname{rank}(X)=1,
\end{aligned}
\end{equation}
where we can identify $X=\ketbra{v}$.

Dropping the non-convex rank constraint $\mathrm{rank}(X)=1$ yields a convex SDP [see also Eqs.~\eqref{eq:trRX} and \eqref{eq:const}]. However, it can be shown that the optimum of this relaxed SDP is attained at the rank-one matrix $X^*=\ketbra{v_{\text{max}}}$, where $\ket{v_{\text{max}}}$ is any unit eigenvector of $R$ associated with the largest eigenvalue $\lambda_{\text{max}}(R)$ \cite{Burer2020Exact,Shor1987Quadratic}. Hence the solution to the task~\eqref{eq:maxFQv} is $\ket{v^*}=2\ket{v_{\text{max}}}$, and the optimal objective value is 
\begin{equation}
\FQ^{(\text{max})}=\braket{v^*}{R|v^*}=4\lambda_{\text{max}}(R). 
\label{eq:maxFQR}
\end{equation}
 
As an illustrative example, let us apply the above procedure to calculate $\FQ^{(\text{max})}[\varrho,\mathcal H]$ for the two-qubit isotropic state $\varrho_{AB}^{0.1,2}$ from Eq.~\eqref{eq:iso} (see also the first line of Table~\ref{tab:QuadraticAndWY}). In this case, the matrix $R$ from Eq.~\eqref{eq:R} evaluates to 
\begin{equation}
R = \frac{162}{95}\,
\begin{bmatrix}
0 & 0 & 0 & 0 & 0 & 0 & 0 & 0\\
0 & 1 & 0 & 0 & 0 & 1 & 0 & 0\\
0 & 0 & 1 & 0 & 0 & 0 & -1 & 0\\
0 & 0 & 0 & 1 & 0 & 0 & 0 & 1\\
0 & 0 & 0 & 0 & 0 & 0 & 0 & 0\\
0 & 1 & 0 & 0 & 0 & 1 & 0 & 0\\
0 & 0 & -1 & 0 & 0 & 0 & 1 & 0\\
0 & 0 & 0 & 1 & 0 & 0 & 0 & 1
\end{bmatrix}.
\end{equation}
The eigenvalues of this symmetric matrix are 
\begin{equation}
\{\lambda_k\}_{k=1}^8=(162/95)\{2,2,2,0,0,0,0,0\}.    
\end{equation}
Hence, by Eq.~\eqref{eq:maxFQR}, we obtain 
\begin{equation}
\FQ^{(\text{max})}[\varrho_{AB}^{0.1,2},\mathcal H]=4\lambda_{\text{max}}(R)=\frac{1296}{95}\simeq 13.6421.
\label{eq:maxFQnum}
\end{equation}
One of the eigenvectors corresponding to the maximal eigenvalue can be written as \be\ket{v_{\text{max}}}=[0,a,0,b,0,a,0,b]^T,\ee where $a\simeq 0.1625$ and $b=\sqrt{\frac{1}{2}-a^2}\simeq0.6882$. Note that since $\lambda_{\text{max}}(R)$ is three-fold degenerate, any unit vector in the corresponding eigenspace gives an optimal $\ket{v_{\text{max}}}$.  

From this choice we obtain $\ket{v^*}=2\ket{v_{\text{max}}}$, and by Eq.~\eqref{eq:H12} the optimal Hamiltonian is given by the bipartite Hamiltonian given in \EQ{eq:bipartiteH}
with
\begin{equation}
H_1 = H_2 = \sqrt{2}\begin{pmatrix}
b & a \\[4pt]
a & -b
\end{pmatrix}.    
\end{equation}
Incidentally, we find $H_1^2=H_2^2=\openone$, which also satisfies the original (non-relaxed) constraint of Eq.~\eqref{eq:cnHconst2}. Thus, in this case the upper bound in \eqref{eq:maxFQnum} is achievable, and the bound is tight. 

\section{Bound entangled state violating the CCNR criterion maximally}
\label{app:BES}

In this Appendix, we present the $4\times4$ bound entangled state for which the
violation of the CCNR criterion given in \EQ{eq:ccnr} is maximal. The state is
\be
\varrho_{\rm CCNR} = \sum_{i=1}^4 p_i \ketbra{\Psi_i}_{AB} \otimes
\varrho_{A'B'}^{(i)},\label{eq:CCNRBES}
\ee
where the probabilities are 
\bea
p_1 &=& p_2 = p_3 = 1/6, \nonumber\\
p_4 &=& 1/2,
\eea
 where the four Bell states are defined as 
 \bea
  \ket{\Phi^\pm}&=&\frac1 {\sqrt{2}} (\ket{00}\pm\ket{11}),\nonumber\\
 \ket{\Psi^\pm}&=&\frac1 {\sqrt{2}} (\ket{01}\pm\ket{10}),
\eea
and the components for $A'$ and $B'$ are given as 
\bea
\varrho_{A'B'}^{(1)} &=& \ketbra{\Psi^+},\nonumber\\
\varrho_{A'B'}^{(2)} &=& \ketbra{\Psi^-},\nonumber\\
\varrho_{A'B'}^{(3)} &=& \ketbra{\Phi^+},\nonumber\\
\varrho_{A'B'}^{(4)} &=& \left(\openone - \ketbra{\Phi^-}\right)/3.
\eea
For the state in \EQ{eq:CCNRBES}, we have
\be
\vert\vert R(\varrho_{\rm CCNR}) \vert\vert_{\rm tr}=1.5.
\ee
Using the method of \SEC{sec:optgain}, we find that the state is not useful metrologically. We created a MATLAB routine that defines $\varrho_{\rm CCNR}$ presented in this paper.  It is part of the QUBIT4MATLAB package \cite{Toth2008QUBIT4MATLAB,QUBIT4MATLAB_actual_note_href}. The routine \verb|BES_CCNR4x4.m| defines the state  given in Eq.~(\ref{eq:CCNRBES}).  We also included a routine that shows the usage of \verb|BES_CCNR4x4.m|. It is called \verb|example_BES_CCNR4x4.m|.

\bibliography{Bibliography2}

\end{document}